\documentstyle[12pt,aps]{revtex}
\input epsf
\begin{document}
\title{
\hfill\parbox[t]{2in}{\rm\small\baselineskip 14pt
{~JLAB-THY-99-08}\vfill~}
\vskip 2cm
Nonresonant Semileptonic Heavy Quark Decay \\ ~}

\author{Nathan Isgur}
\address{Jefferson Lab, 12000 Jefferson Avenue,
Newport News, Virginia, 23606}
\maketitle

\vspace{2.0 cm}
%%%%%%%%%%%%%%%%%%%%%%%%%%%%%%%%%%%%%%%%%%%%%%%%%%%%%%%%%%%%%%%%%%%%
\begin{center}  {\bf Abstract}  \end{center}
%%%%%%%%%%%%%%%%%%%%%%%%%%%%%%%%%%%%%%%%%%%%%%%%%%%%%%%%%%%%%%%%%%%%

\begin{abstract}

		In both the large $N_c$ limit and the valence quark model, semileptonic
decays are dominated by resonant final states.  
Using Bjorken's 
sum rule in an ``unquenched" version of the quark model, I demonstrate
that  in the heavy quark limit
nonresonant final states should also be produced at a significant 
rate.
By calculating the individual
strengths of a large number of exclusive two-body nonresonant channels, 
I show that the total rate for such processes is highly fragmented.  
I also describe
some very substantial duality-violating suppression factors
which reduce the inclusive nonresonant rate to a few percent
of the total semileptonic rate
for the finite quark masses of $\bar B$ decay, and 
comment on the importance of nonresonant decays as testing grounds for very
basic ideas on the structure, strength, and significance of the $q \bar q$
sea and on quark-hadron duality in QCD.
\bigskip\bigskip\bigskip

\end{abstract}
\newpage

%%%%%%%%%%%%%%%%%%%%%%%%%%%%%%%%%%%%%%%%%%%%%%%%%%%%%%%%%%%%%%%%%%
\section {Introduction}
%%%%%%%%%%%%%%%%%%%%%%%%%%%%%%%%%%%%%%%%%%%%%%%%%%%%%%%%%%%%%%%%%%

%%%%%%%%%%%%%%%%%%%%%%%%%%%%%%%%%%%%%%%%%%%%%%%%%%%%%%%%%%%%%%%%%%
\subsection {Background}
%%%%%%%%%%%%%%%%%%%%%%%%%%%%%%%%%%%%%%%%%%%%%%%%%%%%%%%%%%%%%%%%%%

	In both the large $N_c$ limit \cite{LargeNc} and the valence quark model \cite{ISGW,ISGW2}, semileptonic heavy
quark decays are saturated by resonant final states.  In nature this idealization
is broken by light quark pair creation which gives these infinitely narrow resonances
widths and allows nonresonant final states to appear.
While it is clear that $q \bar q$ pairs play
an important role in this and many other phenomena, it is also clear that they remain poorly
understood:

\begin{enumerate}

	\item Although from their widths it is easy to show that hadrons are full of $q\bar q$ pairs, 
meson and baryon spectroscopies are characterized by the valence degrees of freedom.  In
particular there is no evidence for excitations of the $q\bar q$ sea with respect to the valence quarks.

 \item  Even if one were to assume that the $q\bar q$ sea is frozen out of spectroscopy, it
is easy to show that, unless there is a conspiracy between them, meson loop graphs ought 
to destroy the successes of quark model spectroscopy.  For example, the relative shift of the $\rho$ and $a_1$
due to meson loops formed from their dominant decay modes ($\pi \pi$ and $\rho \pi$, respectively)
is hundreds of MeV \cite{GIonV}.

 \item Related to 2), but even more dramatic, is the relative shift of 
pairs of mesons such as the $\rho$ and $\omega$. They
are degenerate in the quark model and in the large $N_c$ limit, as observed in nature, but
meson loop diagrams would lead one to expect them to develop a mass difference of hundreds of
MeV \cite{GIonOZI}.

 \item Given that meson loop diagrams which generate the $q\bar q$ sea are very strong, it is
surprising that the valence quarks seem to dominate low energy current matrix elements \cite{GIonsbars}. The 
extreme interest generated by the proton spin crisis may be attributed to the fact that it
indicates that there are some current matrix elements where the valence quarks do not dominate.

\end{enumerate}

\medskip

		While the questions $q\bar q$ pair creation raises are ubiquitous, heavy quark systems
({\it e.g.}, $Q\bar d$ and $udQ$) are likely to be the most favorable systems in which to find 
answers:  of all systems governed by strong QCD \cite{CloseQuote}, they are arguably the simplest.  Indeed,
many of their properties can be rigorously derived directly from QCD using heavy quark 
symmetry \cite{IWoriginal,IWspec}.  
Their simplicity also makes them more amenable to modelling than other hadrons.  {\it E.g.}, as 
``the hydrogen atoms of QCD"  with the heavy quark defining an origin of coordinates, the
simplest relativistic constituent quark models can  treat $Q \bar d$ systems using the Dirac equation \cite{VanOrden} rather
than a Bethe-Salpeter-type equation.  Because the heavy quark is removed from 
consideration as $m_Q \rightarrow \infty$, such systems offer 
unique opportunities to study the ``brown muck" one chunk
at a time.

	In this paper I expand the quark model treatment 
of the ``brown muck" in heavy meson semileptonic decay from a simple
valence $\bar d$ or $\bar u$ antiquark confined to $Q$ to include the leading effects of $q \bar q$
pair creation. See Fig. 1. Previous studies have examined
low energy pion emission in the context of 
heavy quark chiral perturbation theory \cite{Qxipt}; this work is the first of which I
am aware to address the full array of nonresonant processes. While the results presented here 
will be specific to heavy quark systems, I will draw lessons
from them and suggest experimental consequences of much wider 
interest.

%%%%%%%%%%%%%%%%%%%%%%%%%%%%%%%%%%%%%%%%%%%%%%%%%%%%%%%%%%%%%%%%%%
\subsection {Nonresonant Final States in $\bar B$ and $D$ Semileptonic Decay}
%%%%%%%%%%%%%%%%%%%%%%%%%%%%%%%%%%%%%%%%%%%%%%%%%%%%%%%%%%%%%%%%%%

	For $D$  decays induced by the underlying $c \rightarrow s \bar \ell \nu_{\ell}$ quark decay, 
$D \rightarrow \bar K \bar \ell \nu_{\ell}$ 
and $D \rightarrow \bar K^* \bar \ell \nu_{\ell}$ are clearly dominant, 
accounting for more than $90\%$ of the inclusive 
$D \rightarrow \bar X_s \bar \ell \nu_{\ell}$  semileptonic rate \cite{PDG}.  
For $\bar B$ decays induced by $b \rightarrow  c \ell \bar \nu_{\ell}$, 
$\bar B \rightarrow  D \ell \bar \nu_{\ell}$ and
$\bar B \rightarrow  D^* \ell \bar \nu_{\ell}$ 
account for $64 \pm 6 \% $ of the inclusive 
$\bar B \rightarrow  X_c  \ell \bar \nu_{\ell}$  semileptonic rate \cite{PDG}.
Given the large
energy release in $b \rightarrow c$ versus $c \rightarrow s$, it is not surprising that
in the former
the two ground state resonances of the $s_{\ell}^{\pi_{\ell}}={1 \over 2}^-$ multiplet \cite{IWspec} 
would account for less of the total
semileptonic rate.  Indeed, it has been argued \cite{NIonWolf} by 
assuming duality between the ISGW2 valence quark
model \cite{ISGW2} and QCD-corrected inclusive calculations \cite{inclusives}
that a {\it complete} ISGW2-based calculation
would predict that $20 \pm 6\%$ of the 
$\bar B \rightarrow  X_c  \ell \bar \nu_{\ell}$ rate should go to 
resonant excited states above the $D$ and $D^*$. (The quoted error corresponds
to an estimate of the theoretical uncertainties in the QCD-corrected 
inclusive rate calculations.) 
ISGW2 as published is in contrast not exhaustive: it explicitly computes the rates to the $L=1$ excited states 
with ${s'}_{\ell}^{{\pi'}_{\ell}}={1 \over 2}^+$ and ${s'}_{\ell}^{{\pi'}_{\ell}}={3 \over 2}^+$ 
and to the first radial excitations with ${s'}_{\ell}^{{\pi'}_{\ell}}={1 \over 2}^-$.
These six lowest-lying excitations give 
$8 \pm 1\%$ of the QCD-corrected 
inclusive rate, implying that an additional $12 \pm 6\%$ of the rate of a complete the ISGW2 calculation
should be in yet more highly excited states (both ordinary mesons and hybrids). The ISGW2 model and 
next-to-leading-order QCD can be compared in this way because
both are valence-quark-plus-glue calculations:  $q \bar q$ pairs are ignored in ISGW2 as 
$1 / {N_c}$ corrections and would enter the partonic level inclusive rates only at order $\alpha_s^2$ 
({\it via}
$1 / {N_c}$-suppressed graphs).

\bigskip

%
% format for inserting a picture, complete with line breaks
%
%%%%%%%%%%%%%%%%%%%
\begin{center}
~
\epsfxsize=4.5in  \epsfbox{Qqqddiagrams.ps}
\vspace*{0.1in}
~
\end{center}
%%%%%%%%%%%%%%%%%%%%%%%

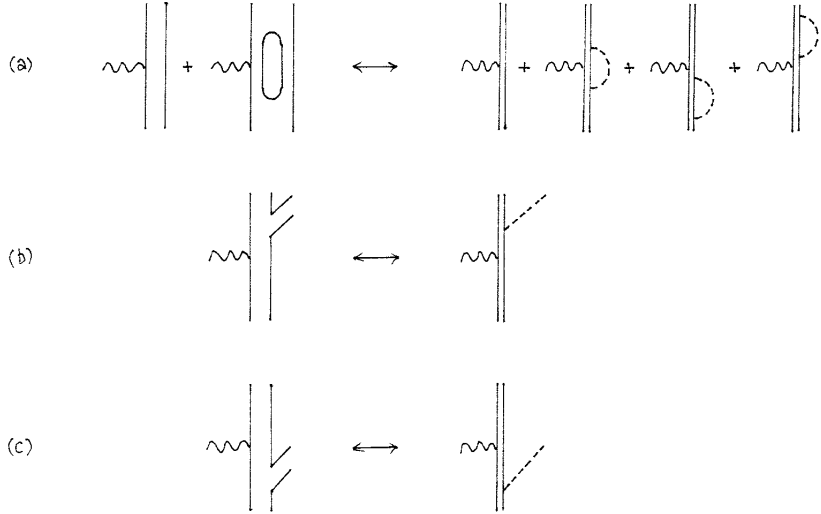
\noindent{ Fig. 1: Leading corrections to narrow resonance saturation of the rates for
semileptonic heavy quark decay, with quark level diagrams on the left and their
hadronic counterparts on the right.  Internal hadronic lines are to be understood as 
summed over their full valence spectra.  a) The $valence \rightarrow valence$ graph
with its leading vertex and external leg corrections, b) The $valence \rightarrow valence$
graph followed by final state decay, and c) Decay from a nonvalence component of the 
initial state to a nonresonant two particle final state.}

\bigskip

	Experimentally, the extent of resonance dominance of $\bar B$ and $D$ semileptonic decays remains
unclear.  In $D$ decay there are explicit measurements giving 
$D \rightarrow \bar K \pi \bar \ell \nu_{\ell}$ 
rates which are $3 \pm 1\%$ of the
semileptonic rate.  Here the $\bar K \pi$
signal is excluded from being the $\bar K^*$, but it is not excluded that it
could arise from the tails of broad resonances, so
even this small fraction cannot be unambiguously identified as nonresonant.  
In $\bar B$ decay, $36 \pm 6\%$ of the
semileptonic rate is not in the $D$ or $D^*$.
Since the dynamical part 
$\rho^2_{dyn} \equiv \rho^2-{1\over 4}$ of the slope of the Isgur-Wise function $\xi(w)$
could be as large as about ${3 \over 4}$ \cite{rho2}, the {\it loss} of rate from these channels
relative to $\rho^2_{dyn}=0$ in the approximation $\xi \simeq 1-\rho^2(w-1)$ 
could indeed be as large as  $36\%$, {\it i.e.,}
the observed non-$(D+D^*)$ rate is not inconsistent with that expected 
from Bjorken's sum rule \cite{Bj,IWonBj} in the heavy
quark limit. Since, as just explained, one expects
$20 \pm 6\%$ of this non-$(D+D^*)$ rate to be in excited resonant states,  
there would be  room for
$16 \pm 8\%$ of $\bar B$ semileptonic decays to be nonresonant,
{\it i.e.}, for a large fraction of $\rho^2_{dyn}$ to be due to
nonresonant channels. There is another closely related indication that
nonresonant channels might be significant: ISGW2 overpredicts the $D$ and $D^*$ rates by amounts which
are consistent with the observation that the Isgur-Wise function is falling about $25\%$ more rapidly
than expected from the opening of just excited resonance decay channels. While
suggestive, both of these observations are also consistent with the ISGW2 model simply underpredicting
decay rates to excited charm states \cite{Wolf}. An additional concern is 
that the suggestive loss of rate from the $D$ and $D^*$ channels calculated from 
Bjorken's sum rule is actually an upper bound in the heavy quark limit: expected quadratic terms
in $\xi$ will dampen this loss, suggesting that perhaps the experimental
non-$(D+D^*)$ fraction is high.

   In summary, there is weak circumstantial evidence for nonresonant
processes in heavy quark semileptonic decays. 
The most compelling case for the existence of such processes at some level is 
nevertheless the simple
theoretical observation that in
the real world we expect very strong
nonvalence $Q\bar q q \bar d$ components in a $Q \bar d$ state and so expect
some inclusive rate to be lost from resonances and transferred to  continua.  In what 
follows I will make these qualitative expectations, previously outlined in Ref. \cite {NIonWolf}, concrete.

%%%%%%%%%%%%%%%%%%%%%%%%%%%%%%%%%%%%%%%%%%%%%%%%%%%%%%%%%%%%%%%%%%
\subsection {The ISGW and ISGW2 Models}
%%%%%%%%%%%%%%%%%%%%%%%%%%%%%%%%%%%%%%%%%%%%%%%%%%%%%%%%%%%%%%%%%%

   In addition to providing a useful  phenomenological guide to semileptonic decays, 
the ISGW model \cite{ISGW} was in many respects a stepping-stone to heavy quark 
symmetry, as it manifested this symmetry
near zero recoil.  
ISGW2 \cite{ISGW2} is an update of ISGW with many new features required by heavy quark symmetry:
it includes
constraints on the relations between form 
factors away from zero recoil and on the slopes of form factors 
near zero recoil \cite{Bj,IWonBj}, it relates
the naive currents of the quark model to the full weak currents
{\it via} the matching conditions of Heavy Quark Effective Theory (HQET) \cite{HQET}, and 
it modifies the {\it ad hoc}  ISGW prescription for connecting quark model form factors
to physical form factors to be consistent with the constraints
of heavy-quark-symmetry-breaking at order $1/m_Q$. Several other improvements
were also made, including the addition of
heavy-quark-symmetry-breaking color magnetic interactions 
to the quark model's dynamics, the
incorporation of relativistic corrections to the axial coupling constants 
(known to be important in the analogous coupling
$g_A$ in neutron beta decay), and
the use of more realistic form factor shapes, based on the measured pion 
form factor. For a more complete discussion of the foundations, strengths, and weaknesses
of such models, see Refs. \cite{ISGW,ISGW2}.

     In this paper I focus  on correcting these models' neglect of 
the nonvalence components of hadrons, but note
that many of the results to be described here are anticipated in the
extensive discussion of Ref. \cite{ISGW} leading to the conclusion that
nonresonant states could be ignored as a first approximation to the dynamics of semileptonic decays.

%%%%%%%%%%%%%%%%%%%%%%%%%%%%%%%%%%%%%%%%%%%%%%%%%%%%%%%%%%%%%%%%%%
\subsection {Unquenching the Quark Model}
%%%%%%%%%%%%%%%%%%%%%%%%%%%%%%%%%%%%%%%%%%%%%%%%%%%%%%%%%%%%%%%%%%

  Some of the key puzzles associated 
with the nature and importance of $q \bar q$ pairs in low 
energy hadron structure were described above. 
These puzzles and potential solutions to them have been extensively discussed in 
a series of papers on ``unquenching" the quark model \cite{GIonV,GIonOZI,GIonsbars}. 
In the following  I
briefly summarize these solutions, since they are the basis for the study presented here.

%%%%%%%%%%%%%%%%%%%%%%%%%%%%%%%%%%%%%%%%%%%%%%%%%%%%%%%%%%%%%%%%%%%%%
\subsubsection{The Origin and Resiliency of Potential Models}
%%%%%%%%%%%%%%%%%%%%%%%%%%%%%%%%%%%%%%%%%%%%%%%%%%%%%%%%%%%%%%%%%%%%%

   A central puzzle in hadron spectroscopy is the apparent absence of low energy
degrees of freedom beyond those which can be attributed to the valence quarks ({\it e.g.},
gluonic or sea quark excitations). Very closely related to this puzzle is the apparent unimportance of
strong meson loop corrections.

   A simple resolution of this 
puzzle has been proposed~\cite{adiabatic}.  
In the  flux tube model~\cite{IsgPat}, 
the quark potential model arises from 
an adiabatic approximation to the gluonic and extra $q \bar q$ degrees of 
freedom embodied in the flux tube.  
This physics has an analog at short distances where perturbation 
theory applies. There $N_f$ types of light $q \bar q$ pairs 
shift (in lowest order) the coefficient of the 
Coulombic potential from
$\alpha_s^{(0)}(Q^2)=\frac{12\pi}{33 {\it ln}(Q^2/\Lambda_0^2)}$ to
$\alpha_s^{(N_f)}(Q^2)=
\frac{12\pi}{(33-2N_f) {\it ln}(Q^2/\Lambda_{N_f}^2)}$,  
the net effect of such pairs 
thus being to produce a {\it new} effective short distance $Q\bar Q$ 
potential. Similarly, when pairs bubble up in the flux tube ({\it i.e.}, 
when the flux tube breaks to create a $Q\bar q$ plus $q\bar Q$ system 
and then ``heals" back to $Q\bar Q$), their net effect is to 
cause a shift $\Delta E_{N_f}(r)$ in the ground state gluonic energy 
which in turn produces a new long-range effective $Q\bar Q$ 
potential.  

It has indeed been shown~\cite{GIonV} that the net long-distance 
effect of the bubbles is to create a 
new string tension $b_{_{N_f}}$ ({\it i.e.}, that the potential 
remains linear).  Since this string tension is to be 
associated with the observed string tension, after renormalization 
{\it pair creation has no effect on the long-distance structure 
of the quark model in the adiabatic approximation}.  Thus 
the net effect of mass shifts from pair creation
is much smaller than one would na\"\i{ve}ly expect from 
the magnitude of typical hadronic widths:  such shifts can only arise from 
nonadiabatic effects \cite{NIonNonadiabatic}.  

It should be  emphasized that no simple truncation of the 
set of all meson loop graphs can reproduce such results: to recover the adiabatic 
approximation requires summing  over 
large towers of $Q\bar q$ plus $q\bar Q$ intermediate states
to saturate their duality with $q \bar q$ loop diagrams which have strength at high energy.

%%%%%%%%%%%%%%%%%%%%%%%%%%%%%%%%%%%%%%%%%%%%%%%%%%%%%%%%%%%%%%%%%%%%
\subsubsection{The Survival of the OZI Rule}
%%%%%%%%%%%%%%%%%%%%%%%%%%%%%%%%%%%%%%%%%%%%%%%%%%%%%%%%%%%%%%%%%%%%

   There is another puzzle of hadronic dynamics which 
is reminiscent of this one:  the success of the OZI 
rule~\cite{OZI}.  
   A generic OZI-violating amplitude $A_{OZI}$ can 
be shown to  vanish like $1/N_c$, and this is often quoted as a rationale for the OZI rule.
  However, there 
are several unsatisfactory features of this 
``solution" to the OZI mixing problem~\cite{LipkinOZI}.  
Consider $\omega$-$\phi$ 
mixing as an example.  This mixing receives a 
contribution from the virtual hadronic loop process 
$\omega \rightarrow K \bar K \rightarrow \phi$, both 
steps of which are OZI-allowed, 
and each of which scales with $N_c$ like 
$\Gamma^{1/2} \sim N_c^{-1/2}$.
The large $N_c$ result that this OZI-violating amplitude
behaves like $N_c^{-1}$ is thus not peculiar to large $N_c$:  
it just arises from ``unitarity" in the sense that the real and 
imaginary parts of a generic hadronic loop diagram will 
have the same dependence on $N_c$. The usual interpretation of the OZI rule in this
case ~-~-~-~ that ``double hairpin graphs" are dramatically suppressed ~-~-~-~ is
untenable in the light of these OZI-allowed loop diagrams. They expose the 
deficiency of the large $N_c$ argument since $A_{OZI} \sim \Gamma$ is {\it not} a good
representation of the  OZI rule.  (Continuing to use $\omega$-$\phi$ mixing as 
an example, we note that $m_\omega - m_\phi$ is numerically comparable to a 
typical hadronic width, so the large $N_c$ result would 
predict an $\omega$-$\phi$ mixing angle of order unity in 
contrast to the 
observed pattern of very weak mixing which implies 
that $A_{OZI} << \Gamma <<m$.)

   Unquenching 
the quark model thus endangers the na\"\i{ve} quark model's agreement 
with the OZI rule.  It has been shown~\cite{GIonOZI} 
how this disaster is naturally averted in the flux tube 
model through a ``miraculous" set of cancellations 
between mesonic loop diagrams consisting of apparently 
unrelated sets of mesons ({\it e.g.}, the $K\bar K$, 
$K\bar K^*+K^*\bar K$, 
and $K^*\bar K^*$ loops tend to strongly cancel against 
loops containing a $K$ or $K^*$ plus one of the four strange 
mesons of the $L=1$ meson nonets).
 Of course the ``miracle" occurs for a good reason:
the sum of {\it all} hadronic loops is dual to a closed $q \bar q$ loop
created and destroyed by a $^3P_0$ operator \cite{3P0mesons,KI}, but
in the closure approximation such an operator cannot create mixing in other
than a scalar channel. It can also be shown \cite{GIonsbars} that
current matrix elements like $\bar s \gamma^{\mu} s$ vanish in this same
approximation.

%%%%%%%%%%%%%%%%%%%%%%%%%%%%%%%%%%%%%%%%%%%%%%%%%%%%%%%%%%%%%%%%%%%%%
\subsubsection{A Summary Comment on Modelling the Effects of $q \bar q$ Pairs}
%%%%%%%%%%%%%%%%%%%%%%%%%%%%%%%%%%%%%%%%%%%%%%%%%%%%%%%%%%%%%%%%%%%%%

    The preceding discussion strongly suggests 
that models which have not addressed the effects of
unquenching on  spectroscopy and the OZI rule should be viewed very 
skeptically as models of the effects of the 
$q \bar q$ sea on hadron structure:
large towers of 
mesonic loops are required to understand how quarkonium 
spectroscopy and the OZI rule survive once strong pair 
creation is turned on.  In particular, while pion and kaon loops (which 
tend to break the closure approximation due to their 
exceptional masses) have a special role to play, 
they will not allow a satisfactory solution to these fundamental
problems associated with unquenching the quark model and so cannot be expected to provide a   
reliable guide to the physics of $q \bar q$  pairs.

    Indeed,  I hope the reader can
appreciate just on the basis of this lightning review that there are
great dangers in drawing conclusions about the strength, structure, or significance
of $q \bar q$ pairs in hadrons from any model that has not dealt with these issues.

%%%%%%%%%%%%%%%%%%%%%%%%%%%%%%%%%%%%%%%%%%%%%%%%%%%%%%%%%%%%%%%%%%
\section {Unquenching Heavy Quark Decay}
%%%%%%%%%%%%%%%%%%%%%%%%%%%%%%%%%%%%%%%%%%%%%%%%%%%%%%%%%%%%%%%%%%

%%%%%%%%%%%%%%%%%%%%%%%%%%%%%%%%%%%%%%%%%%%%%%%%%%%%%%%%%%%%%%%%%%
\subsection {Background}
%%%%%%%%%%%%%%%%%%%%%%%%%%%%%%%%%%%%%%%%%%%%%%%%%%%%%%%%%%%%%%%%%%

	To unquench predictions of the quark model for semileptonic heavy quark decay, I will apply
without alteration the model of Refs. \cite{GIonV,GIonOZI,GIonsbars} which
solves the phenomenological problems associated with unquenching the quark model.  In particular, I 
assume that the $q \bar q$ pair
is created by the action of a pair creation Hamiltonian density $H_{pc}(x)$ in the $Q \bar d$ flux
tube. I further assume that the pair is created with a 
nonlocality (corresponding 
to a finite constituent quark radius) in the coordinate $\vec v$.  See Fig. 2.

\bigskip

%
% format for inserting a picture, complete with line breaks
%
%%%%%%%%%%%%%%%%%%%
\begin{center}
~
\epsfxsize=3.5in  \epsfbox{Qqqdcoords.ps}
\vspace*{0.1in}
~
\end{center}
%%%%%%%%%%%%%%%%%%%%%%%

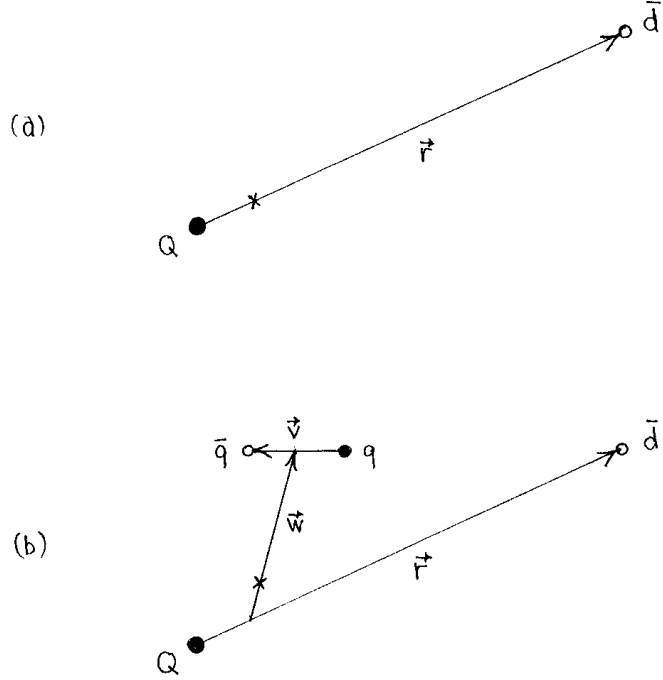
\noindent{ Fig. 2: The coordinates for a) the $Q \bar d$ system, and b) the $Q \bar q q \bar d$ system.
In each
diagram the cross $\times$ denotes the location of the center-of-mass; most results presented in this
paper are in the heavy quark limit where the center-of-mass coincides with the position
of $Q$.}

\bigskip

	The coordinates used here in $Q \bar d$ are the standard center-of-mass and relative coordinates
\begin{eqnarray}
\vec R&=&{{m_Q \vec r_Q+m_d \vec r_{\bar d}} \over m_{Q \bar d}} \simeq \vec r_Q 
\label{eq:defR} \\
\vec r&=&\vec r_{\bar d}-\vec r_Q
\end{eqnarray}
while
in  $Q \bar q q \bar d$ the choice is 
\begin{eqnarray}
\vec R'&=&{{m_Q \vec r_Q+m_q(\vec r_q+\vec r_{\bar q})+m_d \vec r_{\bar d}} \over m_{Q \bar q q\bar d}} \simeq \vec r_Q 
\label{eq:defR'} \\
\vec r&=&\vec r_{\bar d}-\vec r_Q \\
\vec v&=&\vec r_{\bar q}-\vec r_q  \\
\vec w&=&{1 \over 2}(\vec r_q+\vec r_{\bar q})-{{m_Q \vec r_Q+m_d \vec r_{\bar d}} \over m_{Q \bar d}} \simeq
{1 \over 2}(\vec r_q+\vec r_{\bar q})-\vec r_Q~,
\end{eqnarray}
where $m_{ij...k} \equiv m_i+m_j+...+m_k$. 
(Note that the three-vector coordinate $\vec w$ should not be confused with the Lorentz invariant 
heavy quark scalar product
$w=v' \cdot v$ called ``double $u$" (or, in many European
countries, ``double $v$") used in heavy quark form factors). Inverting, we have in $Q \bar d$
\begin{eqnarray}
\vec r_Q&=&\vec R_{cm}-\epsilon_{d/Q\bar d}~ \vec r \simeq \vec R_{cm} \\
\vec r_{\bar d}&=&\vec R_{cm}+\epsilon_{Q/Q\bar d} ~\vec r \simeq \vec R_{cm}+\vec r
\end{eqnarray}
where $\epsilon_{\alpha/\beta} \equiv m_{\alpha}/m_{\beta}$. In
$Q \bar q q \bar d$ we have
\begin{eqnarray}
\vec r_Q&=&\vec R'_{cm}-\epsilon_{q\bar q/Q\bar q q \bar d} ~\vec w -\epsilon_{d/Q\bar d} ~\vec r  \simeq 
\vec R'_{cm} \\
\vec r_{\bar q}&=&\vec R'_{cm}+\epsilon_{Q\bar d/Q\bar q q \bar d} ~\vec w + {1 \over 2} \vec v  \simeq 
\vec R'_{cm}+ \vec w + {1 \over 2} \vec v \\
\vec r_q&=&\vec R'_{cm}+\epsilon_{Q\bar d/Q\bar q q \bar d} ~\vec w - {1 \over 2} \vec v  \simeq 
\vec R'_{cm}+ \vec w - {1 \over 2} \vec v \\
\vec r_{\bar d}&=&\vec R'_{cm}-\epsilon_{q\bar q/Q\bar q q \bar d}~ \vec w +\epsilon_{Q/Q\bar d}~ \vec r \simeq
\vec R'_{cm}+\vec r~.
\end{eqnarray}
Note that most of the results of this paper are presented
in the heavy quark limit where the approximations shown in these formulas will often be used.

    Thus for a $0^-$ state
\begin{equation}
\Psi_{Q \bar d} =  {1 \over {(2 \pi)^{3\over 2}}} e^{i\vec P_{cm}\cdot \vec R} 
\psi_{Q \bar d}(\vec r~) \chi^0_{s_Q s_{\bar d}}
\label{eq:Qbard}
\end{equation}
where $\chi^0_{s_Q s_{\bar d}}$ is the spin zero wavefunction, so that
\begin{equation}
\Phi_{Q \bar d} = \delta^3(\vec P - \vec P_{cm})  
\phi_{Q \bar d}(\vec p) \chi^0_{s_Q s_{\bar d}}
\end{equation}
is the momentum space $Q \bar d$ wavefunction with
\begin{equation}
\phi_{Q \bar d}(\vec p) \equiv {1 \over {(2 \pi)^{3\over 2}}} 
\int d^3r e^{-i\vec p\cdot \vec r} \psi_{Q \bar d}(\vec r~)
\end{equation}
and accordingly
\begin{equation}
\vert P_{Q \bar d}(\vec P_{cm}) \rangle=
\sqrt{2m_{Q \bar d}}
\int d^3p \phi_{Q \bar d}(\vec p~) \chi^0_{s_Q s_{\bar d}}
\vert Q(\epsilon_{Q/Q \bar d}\vec P_{cm}-\vec p, s_Q) 
\bar d(\epsilon_{d/Q \bar d}\vec P_{cm}+\vec p, s_{\bar d}) \rangle~.
\label{eq:defP2}
\end{equation}
Note that in the limit $\vec P_{cm} \rightarrow 0$
relevant here, the factor $\sqrt{2m_{Q \bar d}} \simeq \sqrt{2E_{P_{Q \bar d}}}$ 
is purely conventional as $m_Q \rightarrow \infty$.

    When the flux-tube-breaking  pair creation Hamiltonian
${\bf H}_{pc}^{q \bar q} \equiv 
\int d^3x H_{pc}^{q \bar q}(0, \vec x)$ acts,
\begin{equation}
{\bf H}_{pc}^{q \bar q} \vert P_{Q \bar d}(\vec P_{cm}) \rangle= 
\eta_{q \bar q}\vert P_{Q \bar q q \bar d}(\vec P_{cm}) \rangle
\label{eq:Hpc}
\end{equation}
where, according to the flux tube model,

\begin{enumerate}

	\item  since the $q \bar q$ pair is created in the flux tube,
its center-of-mass is found in a wavefunction $\psi_{ft}(\vec w, \vec r~)$ 
defined by the flux tube spatial profile,

	\item   the {\it internal} wavefunction of the $q \bar q$ pair has $J^{PC}=0^{++}$ and, 
independent of $\vec w$ and  $\vec r$, is of the form 
$\psi^m_{pc}(\vec v) \cdot \chi^{-m}_{s_q s_{\bar q}}$ where 
$\psi^m_{pc}(\vec v)$ has $L=1$, $\chi^{-m}_{s_q s_{\bar q}}$ has $S=1$, and 
$\psi^m \cdot \chi^{-m}  \equiv {1 \over {\sqrt 3}}(\psi^1 \chi^{-1}-\psi^0 \chi^0+\psi^{-1} \chi^1)$, and

	\item  the amplitude to find the $Q \bar d$ subsystem inside the $Q \bar q q \bar d$ system at
relative separation $\vec r$ is identical to that in the ground state, 
namely $\psi_{Q \bar d}(\vec r~)$, since the pair
creation Hamiltonian  density acts locally and instantaneously on the flux tube.

\end{enumerate}

\noindent	In this formulation, $\eta_{q \bar q}$ defines the strength of the 
pair creation, and the normalized state $\vert P_{Q \bar q q \bar d}(\vec P_{cm}) \rangle$
is determined by the wavefunction just described:
\begin{equation}
\Psi_{Q \bar q q \bar d} =  {1 \over {(2 \pi)^{3\over 2}}} e^{i\vec P_{cm}\cdot \vec R'}
\psi_{ft}(\vec w, \vec r~)
\psi^m_{pc}(\vec v) \cdot \chi^{-m}_{s_q s_{\bar q}} 
\psi_{Q \bar d}(\vec r~) \chi^0_{s_Q s_{\bar d}}
\end{equation}
The component parts of this wavefunction and of Eq. (\ref{eq:Qbard})
are defined by previous studies. From ISGW2 \cite{ISGW2},
\begin {equation}
\psi_{Q  \bar d} \simeq {\beta^{3/2}_{Q  \bar d} \over \pi^{3/4}}e^{-{1 \over 2}\beta^2_{Q  \bar d}r^2}
\end {equation}
where $\beta_{Q  \bar d}=0.41$ GeV as $m_Q \rightarrow \infty$ 
as determined variationally from a 
Coulomb-plus-linear-plus-hyperfine Schrodinger equation.  The pair creation wavefunction 
$\vec \psi_{pc}$ is constrained in
Ref. \cite{GIonV} by fitting decay data assuming the form 
\begin {equation}
\vec \psi_{pc} \sim \vec v e^{-3v^2/8r^2_q} \equiv \vec v e^{-{1 \over 2} \beta^2_{pc} v^2}
\label{eq:psipc}
\end {equation}
to have  a quark radius $0 < r_q < 0.4$ fm. 
Given this constraint, I will take $r_q=0.3$ fm as a ``canonical" value, corresponding
to   $\beta_{pc} \simeq 0.58$ GeV, but will consider deviations of $\pm 0.1$ fm from this value as plausible.
(This central value and range are guided by the difficulty of inventing a mechanism
which could lead to a constituent quark radius  $r_q < 0.2$ fm.)
Ideally \cite{KI}, $\psi_{ft}(\vec w, \vec r~)$ should 
have a probability profile which is a tube around the $Q \bar d$ axis
with ``caps" at $Q$ and $\bar d$.  This structure is probably very significant
for the decays of highly excited states, but since all our
decays will emerge from the $Q \bar d$ ground state, I adopt a simpler and more heuristic model which 
simply takes
\begin {equation}
\psi_{ft}(\vec w)={\beta^{3/2}_{ft} \over \pi^{3/4}}e^{-{1 \over 2}\beta^2_{ft}w^2}
\label{eq:psift}
\end {equation}
with $\beta^2_{ft}=fb$, $b$ being the string tension and $f$ a coefficient 
with ``canonical" value $2$ and an uncertainty estimated to be $\pm 1$
based on the calculations of the properties of the ground state wavefunction of
the flux tube
presented in Appendix A of Ref. \cite{KI}. 
Some defects in this simplification and some subtleties associated with both the alternative
of using a flux-tube shape and nonrelativistic kinematics
are discussed in Appendix A of this paper.

	       Since strong decay amplitudes are determined by matrix elements of 
$H^{q \bar q}_{pc}(t, \vec x)$ between
the decaying particle and the continuum, $\eta_{q \bar q}$ can be determined empirically.  In Appendix B, I
extract from $D^* \rightarrow D \pi$ and $K^* \rightarrow K \pi$ 
decays the coefficients $\eta_{q \bar q}$ for $q=u$ or $d$. 
 For concreteness, I will assume following
Refs. \cite{KI,GIonV}, that $\eta_{s \bar s}$ is identical, but 
that the $\eta_{Q \bar Q}$ for $Q=c,b,t$ are all zero.  Finally, for ease
of exposition, I will treat explicitly the case of a single $q \bar q$ flavor in what follows, but 
take into account $q=u,d,s$ in numerical results.

	   Within these approximations,  
\begin{equation}
\Phi_{Q \bar q q \bar d} =  \delta^3(\vec P- \vec P_{cm})
\phi_{ft}(\vec \omega)
\phi^m_{pc}(\vec \pi) \cdot \chi^{-m}_{s_q s_{\bar q}} 
\phi_{Q \bar d}(\vec p~) \chi^0_{s_Q s_{\bar d}}
\end{equation}
where 
$\phi_{ft}$,
$\phi^m_{pc}$, and 
$\phi_{Q \bar d}$
are the momentum space wavefunctions corresponding to 
$\psi_{ft}$,
$\psi^m_{pc}$, and 
$\psi_{Q \bar d}$, respectively,
so that
\begin{eqnarray}
\vert P_{Q \bar q q \bar d}(\vec P_{cm}) \rangle &=&
\sqrt{2m_{Q \bar q q  \bar d}}
\int d^3\omega \int d^3\pi \int d^3p ~\phi_{ft}(\vec \omega)
\phi^m_{pc}(\vec \pi) \cdot \chi^{-m}_{s_q s_{\bar q}} 
\phi_{Q \bar d}(\vec p~) \chi^0_{s_Q s_{\bar d}}  \nonumber \\
&&\vert 
Q(\epsilon_{Q/Q \bar q q \bar d}\vec P_{cm}-\epsilon_{Q/Q  \bar d}\vec \omega - \vec p, s_Q) 
\bar q(\epsilon_{q/Q \bar q q \bar d}\vec P_{cm}+{1 \over 2}\vec \omega + \pi, s_{\bar q})    \nonumber \\
&&
q(\epsilon_{q/Q \bar q q \bar d}\vec P_{cm}+{1 \over 2}\vec \omega - \pi, s_q)
\bar d(\epsilon_{d/Q \bar q q \bar d}\vec P_{cm}-\epsilon_{d/Q  \bar d}\vec \omega + \vec p, s_{\bar d}) 
\rangle~.
\label{eq:defP4}
\end{eqnarray}
(Up to this point, I have retained the exact kinematics of the nonrelativistic
limit for finite $m_Q$, but from now on I will generally 
simplify  results by taking the heavy quark symmetry limit
$m_Q \rightarrow \infty$ with $\vec V_{cm} \equiv \vec P_{cm}/m_Q$ fixed.)

	   While Eq. (\ref{eq:Hpc}) defines the action of ${\bf H^{q \bar q}_{pc}}$ 
on the $Q \bar d$ sector, it does not of course
provide us with the ${\bf H^{q \bar q}_{pc}}$-perturbed ground state.  This state is of the form
\begin{equation}
\vert P_{Q}(\vec P_{cm}) \rangle=
{1 \over {\sqrt{1+c^2_{q \bar q}}}}
\Bigl[ \vert P_{Q \bar d}(\vec P_{cm}) \rangle
+c_{q \bar q}\vert \tilde P_{Q \bar q q \bar d}(\vec P_{cm}) \rangle \Bigr]
\label{eq:mixedstate}
\end{equation}
where $\vert \tilde P_{Q \bar q q \bar d}(\vec P_{cm}) \rangle$ is a normalized 
$Q \bar q q \bar d$ state of the same general form as Eq. (\ref{eq:defP4}) 
but with a wavefunction $\tilde \Phi_{Q \bar q q \bar d}$ to be
specified below and determined via
\begin{eqnarray}
c_{q \bar q}\vert \tilde P_{Q \bar q q \bar d}(\vec P_{cm}) \rangle 
&=&
\sum_{ab} \int {d^3q \over 2m_{Q \bar q q \bar d}}   \nonumber \\
&&{
{\vert \vec P_{cm}; ab(\vec q~)  \rangle
\langle \vec P_{cm}; ab(\vec q~) \vert
{\bf H^{q \bar q}_{pc}}
\vert  P_{Q \bar d}(\vec P_{cm}) \rangle}
\over
{E_{ab}(\vec P_{cm},\vec q~)-E_{P_{Q \bar d}}(\vec P_{cm})}
}~~~.
\label{eq:deftildeP}
\end{eqnarray}
Here $\vert \vec P_{cm}; ab(\vec q~)  \rangle$ is the two meson 
eigenstate with $Q \bar q$ in internal state $a$, $q \bar d$ in internal state $b$, 
and with $(Q \bar q)_a$ and $(q \bar d)_b$
having relative momentum $\vec q$ and total momentum $\vec P_{cm}$; 
$E_{ab}(\vec P_{cm}, \vec q~)$ and $E_{P_{Q \bar d}}(\vec P_{cm})$ are the total energies
of the respective $Q \bar q q \bar d$  and $Q  \bar d$ states
with fixed total center-of-mass momentum $\vec P_{cm}$.  To obtain a rough expression for
$\vert \tilde P_{Q \bar q q \bar d} \rangle$, I make use of the approximate
duality between each of the towers of states $a$ and $b$ and their corresponding 
free particle spectra in the internal
relative momenta $\vec p_{Q\bar q}$ and $\vec p_{q\bar d}$, respectively.  {\it E.g.}, I use
\begin{equation}
\sum_{a} \langle (Q\bar q)_a \vert \simeq \sum_{s_Qs_{\bar q}} 
\int d^3p_{Q\bar q}\langle Q(-\vec p_{Q\bar q}, s_Q)\bar q(\vec p_{Q\bar q},s_{\bar q}) \vert
\label{eq:freeduality} 
\end{equation}
where for simplicity I have illustrated the duality equation in the $Q \bar q$ center of mass frame.
While this replacement is imperfect for low $a$ (corresponding to small $\vec p_{Q\bar q}$), 
since ${\bf H^{q \bar q}_{pc}}$ 
is quite pointlike, Eq. (\ref{eq:deftildeP}) has most
of its strength for relatively massive states $a$ and $b$ and the use of
duality should
be satisfactory for our purposes.

     With these approximations 
we can change  variables from  $a,b, \vec q$ to $\vec p_{Q\bar q}$,  $\vec p_{q\bar d},\vec q$, 
and then convert both sides of
Eq. (\ref{eq:deftildeP}) to  the variables $\vec \omega$,
$\vec \pi$, and  $\vec p$ to identify
\begin {equation}
c_{q \bar q} \tilde \Phi_{Q \bar q q \bar d} \simeq  
{{\eta_{q \bar q}  \Phi_{Q \bar q q \bar d}} \over {\Delta E}}
\label{eq:defcqbarq}
\end {equation}
where in the rest frame $\vec P_{cm}=\vec 0$ and with $m_Q \rightarrow \infty$,
\begin {equation}
\Delta E = 2m_q+{\pi^2 \over m_q}+{\omega^2 \over 4m_q}+\delta
\end {equation}
with 
\begin {equation}
\delta=m_Q+m_d+{p^2 \over 2m_d}-m_{P_{Q\bar d}}~.
\end {equation}
In the duality approximation we have adopted, and with the wavefunction of the $Q \bar d$ 
system identical in $Q \bar q q \bar d$ and $Q  \bar d$, $\delta \simeq 0$.  
Moreover, using the wavefunctions (\ref{eq:psipc}) and (\ref{eq:psift}) and the 
parameters given earlier
\begin {equation}
{ <{ \omega^2 \over 4m_q}> \over   <{ \pi^2 \over m_q}>} 
\simeq
{3 \beta^2_{ft} \over 20 \beta^2_{pc}} 
\end {equation}
is small so that	 
\begin {equation}
\Delta E \simeq 2m_q+{ \pi^2 \over m_q}
\end {equation}
and we may deduce that
\begin{equation}
\tilde \Phi_{Q \bar q q \bar d} \simeq  \delta^3(\vec P- \vec P_{cm})
\phi_{ft}(\vec \omega)
\tilde \phi^m_{pc}(\vec \pi) \cdot \chi^{-m}_{s_q s_{\bar q}} 
\phi_{Q \bar d}(\vec p~) \chi^0_{s_Q s_{\bar d}}
\end{equation}
{\it i.e.}, that $\tilde \Phi_{Q\bar q q \bar d}$ differs from 
$\Phi_{Q\bar q q \bar d}$ simply by the replacement $\phi_{pc} \rightarrow
\tilde \phi_{pc}$ where
\begin{equation}
\tilde \phi^m_{pc}(\vec \pi) \equiv n^{-1/2}
{
{\phi_{pc}^m(\vec \pi)}
\over
{1+\pi^2/2m_q^2}
}~.
\label{eq:phipc}
\end{equation}
The normalization factor $n$ is given by
\begin {equation}
n=\int d^3\pi \vert 
{
{\phi_{pc}^m(\vec \pi)}
\over
{1+\pi^2/2m_q^2}
}
\vert^2~.
\end {equation}
which  may be quite well approximated by the
formula
$n={{1+y}\over{1+21y+12y^3}}$
with 
$y={\beta^2_{pc} \over 8m_q^2}$.
From Eqs. (\ref{eq:defcqbarq}) and (\ref{eq:phipc}) follows the key relation that
\begin {equation}
c_{q \bar q}={
{n^{1/2} \eta_{q \bar q}}
\over
2m_q
}~~~.
\label{eq:candeta}
\end {equation}

    Note that $\tilde \phi^m_{pc}(\vec \pi)$ is softer than $\phi^m_{pc}(\vec \pi)$. We will often use a
harmonic approximation
\begin{equation}
\tilde \phi^m_{pc}(\vec \pi)={{\pi^m} \over \pi^{3/4} \tilde \beta_{pc}^{5/2}}e^{-\pi^2/2 \tilde \beta^2_{pc}}
\label{eq:harmonictildephi}
\end{equation}
to $\tilde \phi^m_{pc}$. It is of course the softer shape of $\tilde \phi^m_{pc}$ that makes $n<1$, and
thus $\tilde \beta_{pc}$ can be determined by requiring that Eqs. (\ref{eq:phipc})
and (\ref{eq:harmonictildephi}) 
match near $\vec \pi=0$,
{\it i.e.}, that 
\begin{equation}
\tilde \beta_{pc}=n^{1/5}  \beta_{pc}~~.
\end{equation}
With realistic parameter values, $n^{1/5} \sim 0.7$, so the softening
is not dramatic: with our canonical value for $\beta_{pc}$, $\tilde \beta_{pc} \simeq 0.4$ GeV.

%%%%%%%%%%%%%%%%%%%%%%%%%%%%%%%%%%%%%%%%%%%%%%%%%%%%%%%%%%%%%%%%%%%%%%%%%%%
\subsection{The Unquenched Isgur-Wise Function}
%%%%%%%%%%%%%%%%%%%%%%%%%%%%%%%%%%%%%%%%%%%%%%%%%%%%%%%%%%%%%%%%%%%%%%%%%%%

     We are now in a position to calculate the unquenched quark model contribution to the Isgur-Wise
function \cite{IWoriginal}.  By heavy quark symmetry,
the form factors for a general $Q_1 \rightarrow Q_2$ transition can be calculated
as matrix elements for the simpler 
$Q \rightarrow Q$ transition with an arbitrary current $\bar Q \Gamma Q$. We therefore focus on the matrix
elements of the scalar current $\bar Q  Q$ between $Q$-containing states: 
\begin {eqnarray}
\xi^{QM}(w)&=&
{1 \over 2m_{Q }}\langle P_Q({\vec P_{cm} \over 2}) \vert \bar Q Q \vert  P_Q(-{\vec P_{cm} \over 2}) \rangle \\
&=&
{1 \over {1+c_{q \bar q}^2}}
[     
\xi_{Q \bar d}^{QM}(w)+c_{q \bar q}^2 \xi_{Q \bar q q \bar d}^{QM}(w)
]
\label{eq:xiqm}
\end {eqnarray}
where
$w \equiv v' \cdot v \simeq 1+P_{cm}^2/2m^2_{Q } = 1+V_{cm}^2/2$ and where,
\begin {eqnarray}
\xi^{QM}_{Q \bar d}(w)&=&
{1 \over 2m_{Q }}\langle P_{Q \bar d}({\vec P_{cm} \over 2}) \vert 
\bar Q  Q\vert  P_{Q \bar d}(-{\vec P_{cm} \over 2}) \rangle \\
&=&
\int d^3r \psi^*_{Q \bar d}(\vec r~)e^{-i m_d \vec V_{cm} \cdot \vec r}\psi_{Q \bar d}(\vec r~)
\label{eq:xi2}
\end {eqnarray}
and
\begin {eqnarray}
\xi^{QM}_{Q \bar q q \bar d}(w)&=&
{1 \over 2m_{Q}}\langle P_{Q \bar q q \bar d}({\vec P_{cm} \over 2}) \vert 
\bar Q Q \vert  P_{Q \bar q q \bar d}(-{\vec P_{cm} \over 2}) \rangle \\
&=&
\xi^{QM}_{Q \bar d}(w) \xi^{QM}_{ft}(w)
\label{eq:xi4}
\end {eqnarray}
with
\begin {equation}
\xi^{QM}_{ft}(w) \equiv
\int d^3w \psi^*_{ft}(\vec w)e^{-2i m_q \vec V_{cm} \cdot 
\vec w}\psi_{ft}(\vec w)~.
\label{eq:xift}
\end {equation}
(In these equations the notation ``$QM$" reminds us that we are calculating in the quark model
so that the perturbative matching of these HQET matrix elements to field theory must be done
at the quark model scale $\mu_{QM} \sim 1$ GeV.)

      {\it We see from  Eq. (\ref{eq:xift}) that $\xi^{QM}_{Q \bar q q \bar d}$ does not depend on the poorly 
known $q \bar q$ wavefunction $\psi_{pc}$}. This simplification
arises because this $q \bar q$ wave function defines the relative position of the $q$ and $\bar q$,
while $\xi^{QM}_{ft}$ is sensitive only to the $q \bar q$ system's wave function relative to $Q$.  
Defining for $w-1 <<1$
\begin {eqnarray}
\xi^{QM} & \simeq & 1-  \rho^2_{wf}(w-1) \\
\xi^{QM}_{Q  \bar d} & \simeq & 1- \rho^2_{Q  \bar d}(w-1) \\
\xi^{QM}_{ft} & \simeq & 1- \rho^2_{ \bar q q  }(w-1)
\end {eqnarray}
and recalling the conventional definition
\begin {equation}
\xi \simeq  1- \rho^2( w-1)
\end {equation}
we therefore have (displaying now explicitly the effects of summing over $q=u$, $d$, and $s$) simply
\begin {equation}
\rho^2_{wf}=
\rho^2_{Q  \bar d}
+{
{\sum_q c^2_{q \bar q}  \rho^2_{  \bar q q  }}
\over
{1+\sum_q c^2_{q \bar q}}
}
\end {equation}
where $\rho^2_{wf}$ is the nonrelativistic wavefunction contribution to $\rho^2$, 
{\it i.e.}, it excludes the relativistic $1 \over 4$
and the contribution $\Delta \rho^2_{pert}$ from matching to the low energy effective theory \cite{ISGW2}. 
Using Eqs. (\ref{eq:xi2}) and (\ref{eq:xi4})
we then have the old result
\begin {equation}
 \rho^2_{Q  \bar d}={m_d^2<r^2> \over 3}={m_d^2 \over 2 \beta^2_{Q \bar d}}
\label{eq:rhoQbard}
\end {equation}
and the new correction from $q \bar q$ pairs
\begin {equation}
 \rho^2_{  \bar q q  }=
{4m_q^2<w^2> \over 3}=
{2m_q^2 \over \beta^2_{ft}}
\end {equation}
so that in the $SU(3)$ limit where $m_u=m_d=m_s=m_q$
\begin {eqnarray}
\rho^2_{wf}&=&
{m_d^2 \over 2 \beta^2_{Q \bar d}}
+{
{\sum_q c^2_{q \bar q} 
{2m_q^2 \over \beta^2_{ft}}}
\over
{1+\sum_q c^2_{q \bar q}}
}
=
{m_d^2 \over 2 \beta^2_{Q \bar d}}
+{
{ c^2_{q \bar q} }
\over
{1+3 c^2_{q \bar q}}
}
\Bigl({6m_q^2 \over \beta^2_{ft}}\Bigr)
 \\
& \equiv &
{m_d^2 \over 2 \beta^2_{Q \bar d}}
+
\Delta \rho^2_{sea} ~.
\label{eq:mainresult}
\end {eqnarray}
This is one of our main new results.  It shows that even if the $c_{q \bar q}^2$ are large, the contribution
of pairs to $\rho^2$ will be small in the adiabatic limit where they are highly localized in the
flux tube ({\it i.e.}, as $\beta^2_{ft} \rightarrow \infty$). See Appendix A for a discussion
of this result for a more general flux-tube shape. We will see next 
that to the extent that pairs  contribute to the
exclusive ``elastic" slope $\rho^2$, they will also contribute to the {\it inclusive} 
nonresonant semileptonic rate.

%%%%%%%%%%%%%%%%%%%%%%%%%%%%%%%%%%%%%%%%%%%%%%%%%%%%%%%%%%%%%%%%%%%%%%%%%%%%%%%%%%%%%
\subsection{A Duality Interpretation of $\Delta \rho^2_{sea}$ via Bjorken's Sum Rule}
%%%%%%%%%%%%%%%%%%%%%%%%%%%%%%%%%%%%%%%%%%%%%%%%%%%%%%%%%%%%%%%%%%%%%%%%%%%%%%%%%%%%%

%%%%%%%%%%%%%%%%%%%%%%%%%%%%%%%%%%%%%%%%%%%%%%%%%%%%%%%%%%%%%%%%%%%%%%%%%%%%%%%%%%%%%
\subsubsection{Motivation}
%%%%%%%%%%%%%%%%%%%%%%%%%%%%%%%%%%%%%%%%%%%%%%%%%%%%%%%%%%%%%%%%%%%%%%%%%%%%%%%%%%%%%

    We have just seen that 
even if $P_{Q_1}$ is full of $q \bar q$ pairs, they may not contribute to $\rho^2$.
We will now see that
it is  incorrect to associate the relative probabilities of
$Q_1 \bar d$ and $Q_1 \bar q q \bar d$ in $P_{Q_1}$ with the resonant and nonresonant parts,
respectively, of $Q_1 \rightarrow Q_2$ semileptonic decay. As heavy quark symmetry requires,
at $w=1$ the $Q_1 \rightarrow Q_2$ transition creates only $P_{Q_2}$ and $V_{Q_2}$ of the
ground state ${s'}_{\ell}^{{\pi '}_{\ell}}= {1 \over 2}^-$ multiplet {\it independent of
the structure of the ``brown muck"}. {\it I.e.}, in this limit the 
$Q_2 \bar d$ and $Q_2 \bar q q \bar d$ components of the hadronic final state, no matter
what their relative strengths, form perfectly into the resonant states $P_{Q_2}$ and $V_{Q_2}$.
For $w-1$ small but nonzero, Bjorken's sum rule \cite{Bj,IWonBj} tells us that the loss of rate 
from the ``elastic" transitions  $P_{Q_1} \rightarrow P_{Q_2}$ and $P_{Q_1} \rightarrow V_{Q_2}$
relative to structureless hadrons with $\rho^2=1/4$ will be exactly compensated
by the production of
${s'}_{\ell}^{{\pi '}_{\ell}}= {1 \over 2}^+$ and ${3 \over 2}^+$
states. In the valence quark model, this rate must appear in $Q_2 \bar d$ excited states.
In Ref. \cite{IWonBj} this valence quark model duality to the quark level semileptonic 
decay  was explicitly demonstrated. Here I will show that the $Q_1 \bar q q \bar d$
content of $P_{Q_1}$ leads in general to the production of both resonant and
nonresonant final states, with the latter rates proportional to $\Delta \rho^2_{sea}$. In
particular, in the adiabatic limit there will be {\it no} nonresonant production.

   To the extent that $\Delta \rho^2_{sea}$ is nonzero, nonresonant $(Q_2 \bar q)_a(q \bar d)_b$
final states with
${s'}_{\ell}^{{\pi '}_{\ell}}= {1 \over 2}^+$ and ${3 \over 2}^+$
will be produced to compensate for the additional loss of rate from the elastic channels
which it causes.
It is natural to expect the compensation to occur in these channels. The
softening of the elastic form factors which depletes the rate to $P_{Q_2}$ and $V_{Q_2}$
will have its analog in inelastic resonance excitation form factors, so these rates will
also be diminished and cannot compensate for the
additional loss of rate from $P_{Q_2}$ and $V_{Q_2}$. 
The population of inelastic channels is thus the only avenue available for
satisfying Bjorken's sum rule. Of course this is also intuitively appealing:
the loss of rate to the elastic channels occurs 
because 
after the recoil from $-\vec P_{cm}/2$ to $+\vec P_{cm}/2$, 
the $q \bar q$ parts of the ground state wave functions of the initial and
final states fail to overlap, and in so doing they must
``by conservation of probability" find themselves in {\it their} excited states, namely as 
$(Q_2 \bar q)_a(q \bar d)_b$ continua. We will now make these heuristic observations precise.

%%%%%%%%%%%%%%%%%%%%%%%%%%%%%%%%%%%%%%%%%%%%%%%%%%%%%%%%%%%%%%%%%%%%%%%%%%%%%%%%%%%%%
\subsubsection{Production of Nonresonant States at Low $w-1$}
%%%%%%%%%%%%%%%%%%%%%%%%%%%%%%%%%%%%%%%%%%%%%%%%%%%%%%%%%%%%%%%%%%%%%%%%%%%%%%%%%%%%%

\bigskip

   The $n^{th}$ valence state perturbed  by ${\bf H_{pc}^{q \bar q}}$ 
(the generalization of Eq. (\ref{eq:mixedstate})) may be written
\begin{equation}
\vert M^{(n)}_Q(\vec P_{cm}) \rangle
=
cos \theta \vert M^{(n)}_{Q \bar d}(\vec P_{cm}) \rangle
+
sin \theta \vert X^{(n00)}_{Q \bar q q \bar d}(\vec P_{cm}) \rangle
\end{equation}
where 
$\vert M^{(n)}_Q\rangle$,
$\vert M^{(n)}_{Q  \bar d}\rangle$, and 
$\vert X^{(n00)}_{Q \bar q q \bar d}\rangle$ are the generalizations of the states
$\vert  P_Q \rangle \equiv \vert M^{(0)}_Q\rangle$, 
$\vert  P_{Q  \bar d}\rangle \equiv \vert M^{(0)}_{Q  \bar d} \rangle$, and
$\vert \tilde P_{Q \bar q q \bar d}\rangle \equiv \vert X^{(000)}_{Q \bar q q \bar d}\rangle$
of Eqs. (\ref{eq:mixedstate}), (\ref{eq:defP2}), and (\ref{eq:deftildeP}), respectively, and where 
under our assumptions that the state of the flux tube is independent of $n$ (the adiabatic approximation)
and that $H_{pc}^{q \bar q}$ does not affect the coordinate $\vec r$, $\theta$ is independent of $n$.
(The rationale for the notation $(n00)$ will become apparent below.) From this expression one can immediately
obtain the generalization of our result for the elastic transition that
\begin{eqnarray}
\xi^{QM}(w)^{n'n} 
&\equiv&{1 \over 2m_Q}
\langle M^{(n')}_Q(+{ \vec P_{cm} \over 2}) \vert \bar Q Q
\vert M^{(n)}_Q(-{ \vec P_{cm} \over 2}) \rangle \\
&=&
\xi^{QM}_{Q \bar d}(w)^{n'n}[cos^2 \theta 
+
sin^2 \theta \xi^{QM}_{ft}(w)]
\end{eqnarray}
where $\xi^{QM}_{Q \bar d}(w)^{n'n}$ is the valence quark model generalization
of the Isgur-Wise function for $n \rightarrow n'$ transitions and
$\xi^{QM}_{ft}(w)$ is exactly the same $q \bar q$  overlap form factor
that appears in the elastic $n=0 \rightarrow n'=0$ transition. Thus the unquenched result
for small $w-1$ is
\begin{equation}
\xi^{QM}(w)^{n'n} 
=
\xi^{QM}_{Q \bar d}(w)^{n'n}[1-sin^2 \theta \rho^2_{  \bar q q  }(w-1)]
\end{equation}
as for the Isgur-Wise function with, once again, 
$sin^2 \theta={{\sum_q c^2_{q \bar q}} \over {1+\sum_q c^2_{q \bar q}}}$.
Since the production of each inelastic resonant channel occurs with
strength proportional to $w-1$ to a positive integral power, the $q \bar q$ modification 
of $\xi^{QM}(w)^{n'n}$ for $n'>0$ has no effect
on the saturation of Bjorken's sum rule for $\xi(w)$ to order $w-1$ since
it produces effects which are at least of order $(w-1)^2$. This is in accord
with the expectations outlined above that the additional
depletion of elastic rate by $q \bar q$ pairs must be 
compensated by the explicit production of a $(Q \bar q)_a(q \bar d)_b$ continuum.

   To see this we must introduce the states in the continuum orthogonal
to $\vert M^{(n)}_Q \rangle$. To this end we
define a complete set of states $\vert X^{(n \alpha \beta)}_{Q \bar q q \bar d}(\vec P_{cm}) \rangle$
in the $Q \bar q q \bar d$ sector. Here $n$,  $\alpha$, and $ \beta $ are excitation 
quantum numbers associated with the $\vec r$, $\vec w$, and $\vec v$ coordinates, respectively.
These states are {\it not} the  eigenstates of this sector in the absence of
${\bf H_{pc}^{q \bar q}}$:
the eigenstates are the  $\vert \vec P_{cm};ab(\vec q~) \rangle$
defined above. However, we can expand
\begin{equation}
\vert \vec P_{cm};ab(\vec q~) \rangle
=
\sum_{n \alpha \beta}
\phi_{ab}^{(n \alpha \beta)}(\vec q~)^*
\vert X^{(n \alpha \beta)}_{Q \bar q q \bar d}(\vec P_{cm}) \rangle~.
\end{equation}
It follows that to lowest order in $\theta$ (or equivalently the pair creation operator)
we can form an orthogonal set of ${\bf H_{pc}^{q \bar q}}$-perturbed states
\begin{eqnarray}
\vert M^{(n)}_Q(\vec P_{cm}) \rangle
& \simeq &
\vert M^{(n)}_{Q\bar d}(\vec P_{cm}) \rangle
+\theta \sum_{ab} \int d^3q  \phi_{ab}^{(n00)}(\vec q~)\vert \vec P_{cm};ab(\vec q~) \rangle \\
\vert X_{ab}(\vec P_{cm}, \vec q~) \rangle 
& \simeq &
\vert \vec P_{cm};ab(\vec q~) \rangle
-\theta \sum_n  \phi_{ab}^{(n00)}(\vec q~)^* \vert M^{(n)}_{Q\bar d}(\vec P_{cm}) \rangle
\end{eqnarray}
where $(\alpha, \beta)=(0,0)$ define the universal state of
the $q \bar q$ pair created in a flux tube by the action
of ${\bf H_{pc}^{q \bar q}}$. Let us now compute the transition
amplitude to the continuum:
\begin{eqnarray}
\langle X_{ab}(+{\vec P_{cm} \over 2}, \vec q~) \vert \bar Q  Q 
\vert M^{(0)}_Q(-{\vec P_{cm} \over 2}) \rangle
& \simeq &
\theta[
\langle +{\vec P_{cm} \over 2};ab(\vec q~) \vert \bar Q  Q 
\vert \tilde P_{Q \bar q q \bar d} (-{\vec P_{cm} \over 2}) \rangle  \nonumber \\
~~~~~&-&
\sum_n \phi_{ab}^{(n00)}(\vec q~)
\langle M^{(n)}_{Q \bar d}(+{\vec P_{cm} \over 2}) \vert \bar Q Q 
\vert M^{(0)}_{Q \bar d}(-{\vec P_{cm} \over 2}) \rangle]
\\
&& \nonumber \\
& \simeq &
\theta[\sum_{n \alpha \beta} \phi_{ab}^{(n\alpha \beta)}(\vec q~) 
\langle X^{(n\alpha \beta)}_{Q \bar q q \bar d}(+{\vec P_{cm} \over 2}) \vert \bar Q Q 
\vert X^{(000)}_{Q \bar q q \bar d} (-{\vec P_{cm} \over 2}) \rangle   \nonumber \\
~~~~~&-&
\sum_n \phi_{ab}^{(n00)}(\vec q~)
\langle M^{(n)}_{Q \bar d}(+{\vec P_{cm} \over 2}) \vert \bar Q Q 
\vert M^{(0)}_{Q \bar d}(-{\vec P_{cm} \over 2}) \rangle]~.
\label{eq:trans1}
\end{eqnarray}
As $m_Q \rightarrow \infty$, $ \vec P_{cm} = m_Q  \vec V_{cm} $ is much larger
than any internal momentum so the matrix elements
$\langle Q \vert \bar Q Q \vert Q \rangle$
appearing in the $Q \bar d \rightarrow Q \bar d$ and
$Q \bar q  q \bar d \rightarrow Q \bar q q \bar d$ transitions here are identical. Moreover, since the
$\bar Q Q$ current does not affect the internal state of the $q \bar q$ pair,
$\beta$ is required to be zero.
Thus Eq. (\ref{eq:trans1}) becomes
\begin{equation}
\langle X_{ab}(+{\vec P_{cm} \over 2}, \vec q~) \vert \bar Q Q 
\vert M^{(0)}_Q(-{\vec P_{cm} \over 2}) \rangle
\simeq 
\theta \sum_n \xi^{QM}_{Q \bar d}(w)^{n0}
\Bigl[\sum_{\alpha}
\xi^{QM}_{ft}(w)^{\alpha 0}
\phi_{ab}^{(n \alpha 0)}(\vec q~)
-
\phi_{ab}^{(n 0 0)}(\vec q~)\Bigr]
\label{eq:trans2}
\end{equation}
where $\xi^{QM}_{ft}(w)^{\alpha 0}$ is the generalization of $\xi^{QM}_{ft}(w)$
encountered above, namely
\begin{equation}
\xi^{QM}_{ft}(w)^{\alpha 0}
\equiv
\int d^3w~
\psi^{(\alpha)}_{ft}(\vec w)^*e^{-2i m_q \vec V_{cm} \cdot \vec w}
\psi^{(0)}_{ft}(\vec w)
\end{equation}
where $\psi^{(\alpha)}_{ft}$ is the $\alpha^{th}$ basis state for the expansion of
the $q \bar q$ center of mass coordinate $\vec w$. Expanding the exponential in powers of $\vec V_{cm}$
as is appropriate for small $w-1$, we obtain Eq. (\ref{eq:xift}) of Section II for 
$\alpha=0$ and 
\begin{equation}
\xi^{QM}_{ft}(w)^{\alpha \neq 0, 0}
\simeq -2i m_q \vec V_{cm} \cdot 
\int d^3w~
\psi^{(\alpha)}_{ft}(\vec w)^* \vec w
\psi^{(0)}_{ft}(\vec w)~.
\end{equation}
Since $\xi^{QM}_{ft}(w)^{0 0} \simeq 1 - \rho^2_{ \bar q q }(w-1)$
with $w-1=V^2_{cm}/2$, to leading order in $V_{cm}$ the $\alpha=0$ term
in Eq. (\ref{eq:trans2}) cancels with $\phi_{ab}^{(n00)}(\vec q~)$ to leave 
\begin{equation}
\langle X_{ab}(+{\vec P_{cm} \over 2}, \vec q~) \vert \bar Q Q 
\vert M^{(0)}_Q(-{\vec P_{cm} \over 2}) \rangle
\simeq 
\theta \sum_{n, \alpha \neq 0} \xi^{QM}_{Q \bar d}(w)^{n0}
\xi^{QM}_{ft}(w)^{\alpha 0}
\phi_{ab}^{(n \alpha 0)}(\vec q~)~.
\label{eq:trans3}
\end{equation}
An immediate consequence of this relation is that in the adiabatic limit ($\beta_{ft} \rightarrow \infty$),
$\xi^{QM}_{ft}(w)^{\alpha 0} \sim \beta_{ft}^{-1} \rightarrow 0$ for
$\alpha \neq 0$ so we have explicitly demonstrated that there is no nonresonant production
in this limit.

    Since we have for our discussion
assumed that $\psi^{(0)}_{ft}(\vec w)$ has the form of a ground state
harmonic oscillator wave function, it is natural to use a harmonic oscillator basis
as the orthonormal expansion functions for the variable $\vec w$. Doing so, it follows
that only three basis states give nonzero contributions to Eq. (\ref{eq:trans3}) to
leading order in $\vec V_{cm}$ since $\vec w \psi^{(0)}_{ft}(\vec w)$ is proportional
to the three $n_w=0$, $\ell_w=1$ harmonic oscillator
wave functions:
\begin{equation}
\psi^{[n_w=0, \ell_w=1,i]}_{ft}(\vec w)
=
\sqrt{2}{\beta_{ft}^{5/2} \over \pi^{3/4}} w^i
e^{-{1 \over 2}\beta_{ft}^2w^2}
\end{equation}
in a Cartesian basis, giving
\begin{equation}
\xi^{QM}_{ft}(\vec w)^{[n_w=0, \ell_w=1,i]0}
=
-{
{i  \sqrt{2} m_q V^i_{cm}} \over
{\beta_{ft}}
}
\end{equation}
and thence
\begin{equation}
\langle X_{ab}(+{\vec P_{cm} \over 2}, \vec q~) \vert \bar Q Q 
\vert M^{(0)}_Q(-{\vec P_{cm} \over 2}) \rangle
\simeq 
\theta \sum_{n, i} \xi^{QM}_{Q \bar d}(w)^{n0}
\xi^{QM}_{ft}(\vec w)^{[n_w=0, \ell_w=1,i]0}
\phi_{ab}^{(n [n_w=0, \ell_w=1,i] 0)}(\vec q~)~.
\end{equation}
Next we note that
\begin{equation}
\xi^{QM}_{Q \bar d}(w)^{n\neq0,0} \sim (w-1)^k
\end{equation}
where $k$ is a positive integer by Luke's Theorem \cite{Luke} so that to order $\vec V_{cm}$
\begin{eqnarray}
\langle X_{ab}(+{\vec P_{cm} \over 2}, \vec q~) \vert \bar Q Q 
\vert M^{(0)}_Q(-{\vec P_{cm} \over 2}) \rangle
& \simeq &
\theta \xi^{QM}_{Q \bar d}(w) \sum_i 
\xi^{QM}_{ft}(\vec w)^{[n_w=0, \ell_w=1,i]0}
\phi_{ab}^{(0 [n_w=0, \ell_w=1,i] 0)}(\vec q~) \\
& \simeq &
-{
{i \theta \sqrt{2} m_q \vec V_{cm}} \over
{\beta_{ft}}
} \cdot
\vec \phi_{ab}(\vec q~)
\end{eqnarray}
where I have introduced the notation $\vec \phi_{ab}(\vec q~)$ for the three-vector
$\phi_{ab}^{(0 [n_w=0, \ell_w=1,i] 0)}(\vec q~)$. 

   This key result has a simple interpretation. Nonresonant production at
order $w-1$ requires that the current $\bar Q Q $ acting on $Q \bar q q \bar d$ create neither the
ground state nor the $P$-wave resonances. However,
it cannot excite the $q \bar q$ internal coordinate $\vec v$ and,
if it excites $\vec r$ then to order $w-1$ it has just produced the $Q \bar q q \bar d$ component
of either the ground state or the $P$-wave resonances. {\it Hence nonresonant 
production to order $w-1$ occurs purely by excitation of the $q \bar q$
coordinate $\vec w$ to $\ell_w=1$.} The
factors $\vec \phi_{ab}(\vec q~)$ are simply the projections of
these $\ell_w=1$ states onto the continuum eigenstates 
consisting of mesons $a$ and $b$ with relative momentum $\vec q$.

   We are now in a position to verify that the $\Delta \rho^2_{sea}$ contribution
to the slope of the Isgur-Wise function is indeed compensated by the production
of these continuum $(Q \bar q)_a (q \bar d)_b$ states. The probability for their production
at $w$ is, up to $(w-1)^2$ corrections,
\begin{equation}
dP(P_Q \rightarrow {\rm{continuum}})
\simeq 
\sum_{ab} \int d^3q
\vert \langle X_{ab}(+{\vec P_{cm} \over 2}, \vec q~) \vert \bar Q Q 
\vert M^{(0)}_Q(-{\vec P_{cm} \over 2}) \rangle \vert^2 
\end{equation}
which since
$
\sum_{ab} \int d^3q
\phi^i_{ab}(\vec q~) \phi^j_{ab}(\vec q~)=\delta^{ij}~,
$
gives
\begin{eqnarray}
dP(P_Q \rightarrow {\rm{continuum}})
& \simeq &
{ 
{2 \theta^2 m_q^2 V^2_{cm} }
\over
\beta^2_{ft}
} \\
& \simeq & {4m_q^2 \over \beta^2_{ft}} \theta^2(w-1)
\label{eq:nrtotal}
\end{eqnarray}
for the flavor $q$. On the other hand, according to Eq. (\ref{eq:mainresult}), the contribution
of flavor $q$ to the loss of elastic rate is
\begin{equation}
dP(P_Q \rightarrow P_Q+V_Q)_{Q \bar q q \bar d} \simeq -{4m_q^2 \over \beta^2_{ft}} \theta^2(w-1)~.
\end{equation}
The two rates match, explicitly demonstrating the connection of $\Delta \rho^2_{sea}$ to
the nonresonant continuum.

%%%%%%%%%%%%%%%%%%%%%%%%%%%%%%%%%%%%%%%%%%%%%%%%%%%%%%%%%%%%%%%%%%%%%%%%%%%%%%%%%%%%%
\subsubsection{Production of Exclusive  Nonresonant States at Low $w-1$}
%%%%%%%%%%%%%%%%%%%%%%%%%%%%%%%%%%%%%%%%%%%%%%%%%%%%%%%%%%%%%%%%%%%%%%%%%%%%%%%%%%%%%

\bigskip

   It remains to assess the fractional population of individual continuum channels inside
of the total given by Eq. (\ref{eq:nrtotal}). To do this we must calculate
$\vec \phi_{ab}(\vec q~)$, which from its definition is
\begin{equation}
\phi_{ab}^{(n \alpha \beta)}(\vec q~) \delta^3(\vec P~'_{cm}-\vec P_{cm})=
\langle \vec P~'_{cm};ab(\vec q~) 
\vert X^{(n \alpha \beta)}_{Q \bar q q \bar d}(\vec P_{cm}) \rangle
\end{equation}
for the case $(n \alpha \beta)=(0 [n_w=0, \ell_w=1,i] 0)$.
These calculations are straightforward, but would be quite tedious without the introduction
of several tricks described  in Appendix C. Results of the
calculations of $\tau^{(m)}_{1/2}(w)$ 
and $\tau^{(p)}_{3/2}(w)$ for  number of
low-lying nonresonant channels are given in Table I.

%%%%%%%%%%%%%%%%%%%%%%%%%%%%%%%%%%%%%%%%%%%%%%%%%%%%%%%%%%%%%%%%%%%%%%%%%%%%%%%%%%%%%
\section{Unquenching Heavy Quark Decay: Results}
%%%%%%%%%%%%%%%%%%%%%%%%%%%%%%%%%%%%%%%%%%%%%%%%%%%%%%%%%%%%%%%%%%%%%%%%%%%%%%%%%%%%%

\bigskip

    We now turn to the quantitative evaluation of
$\Delta \rho^2_{sea}$ (which is a reflection of total nonresonant production) and then to the
distribution of these decays into exclusive nonresonant channels.

%%%%%%%%%%%%%%%%%%%%%%%%%%%%%%%%%%%%%%%%%%%%%%%%%%%%%%%%%%%%%%%%%%%%%%%%%%%%%%%%%%%%%
\subsection{The Total Nonresonant Rate}
%%%%%%%%%%%%%%%%%%%%%%%%%%%%%%%%%%%%%%%%%%%%%%%%%%%%%%%%%%%%%%%%%%%%%%%%%%%%%%%%%%%%%

\bigskip

   As shown in Appendix B, the light quark amplitudes
$\eta_{u \bar u}=\eta_{d \bar d}$
can be determined from strong decays to be
\begin{equation}
\eta_{q \bar q} \simeq 0.9 ~{\rm GeV}
\end{equation}
with an uncertainty of a factor of two mainly arising from a strong
dependence on the poorly known quantity $\beta_{pc}$. It follows from Eq. (\ref{eq:candeta}) that
\begin{equation}
c_{q \bar q} \simeq 0.5~.
\end{equation}
Assuming $SU(3)$
symmetry for the contribution to
$\Delta \rho^2_{sea}$, one then obtains from Eq. (\ref{eq:mainresult}) 
\begin{equation}
\Delta \rho^2_{sea} \simeq {1 \over 4}~,
\end{equation}
corresponding to an increase of $\rho^2$ from the value $0.74 \pm 0.05$ quoted in 
ISGW2 to a value near unity. Either value would be in reasonable agreement with measurements
\cite{rho2}. {\it Via} the Bjorken sum rule, such a $\rho^2_{sea}$ would be consistent with the 
possibility discussed in the Introduction that $16 \pm 8 \%$ of the inclusive semileptonic
$\bar B$ rate is in nonresonant channels.

   This reasonable quantitative correspondence between our calculated
$\Delta \rho^2_{sea}$
and a possible experimental anomaly should not be taken too seriously.  The 
missing non-$(D+D^*)$ rate attributed to nonresonant production might be due
to an ISGW2 underestimate of excited resonance production \cite{Wolf}, or to an
experimental overestimate of non-$(D+D^*)$ production. Moreover, while it is
a ``canonical estimate", $\Delta \rho^2_{sea}$ is subject to very
substantial uncertainties: see Table II.

%%%%%%%%%%%%%%%%%%%%%%%%%%%%%%%%%%%%%%%%%%%%%%%%%%%%%%%%%%%%%%%%%%%%%%%%%%%%%%%%%%%%%
\subsection{The Rate to Low-Lying Exclusive Nonresonant Channels}
%%%%%%%%%%%%%%%%%%%%%%%%%%%%%%%%%%%%%%%%%%%%%%%%%%%%%%%%%%%%%%%%%%%%%%%%%%%%%%%%%%%%%

\bigskip

   Using the amplitudes of Table I and the formulas of Appendix C, one can easily
calculate the {\it fractions} of  $\Delta \rho^2_{sea}$ due to the individual low-mass
nonresonant channels shown in Table III. Note that the nonresonant rate is highly fragmented:
none of the many channels tabulated account for more than $4\%$ of the inclusive
nonresonant rate (and thus no more than about $1\%$ of the total semileptonic rate). 
These results are consistent with previous studies
of single low energy pion emission using heavy quark chiral perturbation theory \cite{Qxipt}.
Note also
that the thirty final states considered here account for only about $40\%$ of the nonresonant rate,
so that most of this rate resides in states $ab$ composed of highly excited resonances. 

   There are several reasons why the results given in
Table III must be interpreted with caution. The most prominent is simply
that they depend roughly on $\tilde \beta_{pc}^{-5}$!
Recalling that $\tilde \beta_{pc}=n^{1/5}  \beta_{pc}$, our canonical value for
$\tilde \beta_{pc}$ is $0.4$ GeV, but over the range of $\beta_{pc}$ allowed
by Table II, it varies from its canonical value by $\pm 0.1$ GeV. Although the factor $n^{1/5}$ makes
this range narrower than that of $\beta_{pc}$ itself, it still leaves us 
with an uncertainty of more than a factor of two on the basic unit
for production of exclusive nonresonant channels.

    The second word of caution concerns large corrections to the $m_Q \rightarrow \infty$ limit
studied here which arise for real $\bar B$ decay. The problem is phase space: in the heavy quark limit,
all final states occur at the $w$ of the underlying quark production process, {\it i.e.}, the
full mass range of the final state spectrum of states is negligible compared to the
heavy quark energies even at low $w-1$. In actual $\bar B \rightarrow X_c \ell \bar \nu_{\ell}$ decays,
each final state $X_c$ will have a Dalitz plot that is a shrinking fraction of the
$b \rightarrow c \ell \bar \nu_{\ell}$ Dalitz plot as $m_{X_c} \rightarrow m_B$. Thus in the
heavy quark limit the loss of rate from the elastic channels $D+D^*$ would be {\it locally
compensated} in the variable $w$. For finite $m_b$ and $m_c$, however, a loss of elastic rate will still
occur {\it via} $\Delta \rho^2_{sea}$, but the compensating channels will experience a delayed
turn-on because of their thresholds; indeed, some processes which would have helped to compensate
$\Delta \rho^2_{sea}$ will be kinematically forbidden. These 
phase space suppressions lead to $\Lambda_{QCD}/m_Q$-type corrections to the inclusive
rate, and therefore also corrections to the accuracy of quark-hadron duality. However, it is
known on general grounds from heavy quark symmetry and the operator product expansion (OPE) 
that, as the energy release $m_b-m_c \rightarrow \infty$, the leading
corrections to the inclusive
rate must be of order $\Lambda^2_{QCD}/m^2_Q$ \cite{inclorig,SVonInclusives}.
The resolution of this apparent paradox has been discussed in Ref. \cite{NIonInclusives}:
the OPE has a radius of convergence which does not include the region
in which (significant) hadronic thresholds are turning on, so for real $b \rightarrow c$ transitions
there can be $\Lambda_{QCD}/m_Q$ corrections, and they can be substantial. However,
associated with these $\Lambda_{QCD}/m_Q$ threshold effects, which would diminish the
integrated contribution of individual hadronic channels below the level required for perfect duality, are
$\Lambda_{QCD}/m_Q$ corrections to the rate for the production of such channels
which {\it enhance} their production once they are above threshold. These counterbalancing effects soften
the breaking of duality: they are the precursors of the perfect cancellation of $\Lambda_{QCD}/m_Q$
effects that must occur as $m_b-m_c \rightarrow \infty$.

    How then should one make use of Table III for real $b \rightarrow c$ transitions? The amplitudes 
 $\tau_{1/2}^{(m)}$ and $\tau_{3/2}^{(p)}$ shown are the leading order predictions
for their respective channels. For the reasons just outlined, we can expect that they will be enhanced
by  $\Lambda_{QCD}/m_Q$ corrections which are expected to be of the order of $25-50\%$ \cite{NIonInclusives}.
However, despite this effect, I believe that the dramatically
reduced size of their Dalitz plots (relative to that for the underlying 
quark process $b \rightarrow c \ell \bar \nu_{\ell}$)
will reduce the actual population of all nonresonant states well below that expected
from $\Delta \rho^2_{sea}$. Indeed, each channel shown in Table III has a continuous 
spectrum of masses from its threshold up to masses exceeding $m_B$. Consider, for example,
the $D+\rho$ channel. Its mass is given by $m_{D \rho}(q^2)={\sqrt {m_D^2+q^2}}+{\sqrt {m_{\rho}^2+q^2}}$,
while its dominant $D$-wave production rate
is proportional to $q^4 e^{-q^2/4 \tilde{ \bar \beta}^2}$, where
$2 \tilde{ \bar \beta} \sim 1$ GeV. The production is therefore very weak at low masses where
the available phase space is generous, and peaks at $m_{D \rho} \sim m_B$ where it vanishes.
Given these basic kinematic facts, it is clear that even this simple exclusive
channel will be produced at a rate far less than in the heavy quark limit, and that most
of $\Delta \rho^2_{sea}$ will be uncompensated. The third column of Table III gives 
the phase space factors by which individual channels will be reduced in real
$\bar B \rightarrow X_c \ell \bar \nu_{\ell}$ decays relative to the heavy quark limit;
the fourth column gives a very rough estimate for the {\it net} suppression of each channel
as a product of this phase space factor and a generous guess that, 
after $\Lambda_{QCD}/m_Q$ corrections, each channel 
has a compensatory increase of $50\%$. Considering that in general
the untabulated channels of yet more highly excited states $ab$ will suffer even greater phase space suppressions, an overall
diminution of the nonresonant rate by at least a factor of four seems likely.

%%%%%%%%%%%%%%%%%%%%%%%%%%%%%%%%%%%%%%%%%%%%%%%%%%%%%%%%%%%%%%%%%%%%%%%%%%%%%%%%%%%%%
\section{Conclusions}
%%%%%%%%%%%%%%%%%%%%%%%%%%%%%%%%%%%%%%%%%%%%%%%%%%%%%%%%%%%%%%%%%%%%%%%%%%%%%%%%%%%%%

\bigskip

   Although the successes of the valence quark model and the arguments of the large $N_c$ limit
provide indications that sea quarks play a relatively minor role
in hadronic physics, this hope is far from being justified by our current
understanding. Some failures of the quark model
({\it e.g.}, the proton spin crisis) and the known existence of strong real and
virtual decay channel couplings indeed make blithely ignoring the role of $q \bar q$ pairs
both phenomenologically and theoretically untenable.
In this work I have examined the influence of $q \bar q$ pairs on the simplest ``real" 
hadrons: heavy quark mesons like the $\bar B$.

    This study has led to a number of qualitative insights which I believe
are quite general in nature.
In earlier work on ``unquenching the quark model", the success of the valence
quark model in spectroscopy was shown to have a possible basis in the validity of an adiabatic
approximation. In this approximation, both the confining
flux tube and the many $q \bar q$ pairs it generates remain in their adiabatically
evolving
ground state as the valence quarks move. In this work I have shown that the same approximation
leads to valence quark and therefore resonance dominance of the simplest current matrix
elements: $\bar Q_2 \Gamma Q_1$ matrix elements of heavy quark mesons. The physical picture behind
our results is simple and appealing. According to heavy quark symmetry, at small recoil
the $Q_1 \rightarrow Q_2$ decay of the $Q_1$ ground state $P_{Q_1}$ will lead with unit probability
to $P_{Q_2}$ and $V_{Q_2}$ {\it no matter how complicated the QCD
``brown muck" might be}. This simple observation makes it clear that for
$\bar Q_2 \Gamma Q_1$ matrix elements the issue is not the probability of
$q \bar q$ pairs in $P_{Q_1}$ but rather how rapidly as $w-1$ increases these pairs
fail to overlap with those in $P_{Q_2}$, $V_{Q_2}$, and the $Q_2 \bar d$ excited states.
I have shown explicitly that in the adiabatic limit these overlaps are perfect so
that only valence states (the resonances) are produced. Moreover, 
I showed that violations of the adiabatic approximation
can be directly associated with the production of nonresonant states. Thus this study
leads to a way of understanding how the valence quark model can be so successful {\it even though}
hadrons are full of $q \bar q$ pairs.

     While they are quantitatively very crude, these calculations also have interesting
consequences for real $\bar B$ semileptonic decays. First of all, they suggest that
$\rho^2_{dyn} \equiv \rho^2-{1 \over 4}$ is composed of two comparable
parts: $\rho^2_{resonant} \simeq {1 \over 2}$ and
$\rho^2_{nonresonant}  \simeq {1 \over 4}$, though we have stressed that this
split of $\rho^2_{dyn}$ is very model-dependent. In the heavy quark limit,
Bjorken's sum rule would then lead one to expect (using the central experimental value of the
$(D+D^*)$ fraction) that roughly $24\%$ (with a $50\%$ error) of semileptonic decays
go into resonances (both ordinary $Q \bar d$ mesons and $Q \bar d$ hybrids), and
roughly $12\%$ (with an error of a factor of two) go into nonresonant states. Since the former states
will have most of their strength in the $2.4-3.0$ GeV region, they will suffer,
relative to the heavy quark limit, phase space suppression
factors varying from only $0.75$ to $0.50$ over this range which may be fully  compensated by the
$\Lambda_{QCD}/m_Q$ enhancements required by asymptotic duality \cite{SVonInclusives,NIonInclusives}.
In contrast, we have seen that nonresonant states are expected to populate very high masses
peaking in strength near $m_B$ and so to suffer a substantial {\it net} suppression factor
of at least ${1 \over 4}$. From this study I therefore expect that ${\buildrel < \over \sim} 5\%$ of
$\bar B$ semileptonic decays will be nonresonant!

    As a corollary of this last observation, I note that if a $12\%$ nonresonant semileptonic 
fraction is required for duality
but only a quarter of this is realized, then duality will fail from this effect alone by $\sim 10\%$,
as anticipated in Ref. \cite{NIonInclusives}. There is, however, a minor inconsistency associated with
this conclusion. Experiment requires that $36 \pm 6\%$ of $\bar B$ semileptonic decays
be non-$(D+D^*)$ decays, in contrast to the $\sim 25\%$ we would have estimated from the preceeding.
As mentioned earlier, this could simply mean that the ISGW2 model underpredicts
the production of excited charm mesons \cite{Wolf} or that experiment has overestimated non-$(D+D^*)$ production.
Determining whether this discrepancy is real will require a more quantitative
calculation than this one (and probably additional experimental measurements as well).

    Detailed experimental studies of the structure of the hadronic final state in 
semileptonic $b \rightarrow c$ decays can therefore answer some fundamental questions about 
the role of $q \bar q$ pairs and about duality in strong QCD. A vital feature of these systems
is that duality is underwritten by
Bjorken's sum rule, requiring an exact and local duality 
between quark- and hadronic-level decays in the heavy quark limit. In particular, 
the experimental determination of the strength and structure of these nonresonant
contributions would immediately test the conclusions reached here that these $q \bar q$ effects
are highly suppressed in real $b \rightarrow c$ decays, that such decays extend to very high masses,
and that they are highly fragmented into many small channels. Independent of the outcome,
examining this problem in Nature's
simplest hadronic system under the action of its simplest
current (a heavy-to-heavy nonsinglet transition) should prove to be an excellent starting point for 
eventually understanding the $q \bar q$ sea  in all strongly interacting matter. In 
particular, given the complexity of QCD, this seems an essential first step before tackling
the problems of duality and nonresonant production in ordinary deep inelastic scattering.

\vfill\eject

%%%%%%%%%%%%%%%%%%%%%%%%%%%%%%%%%%%%%%%%%%%%%%%%%%%%%%%%%%%%%%%%%%%%%%%%%%
{\noindent \bf APPENDIX A: Flux Tubes and A Critique of Nonrelativistic Kinematics}
%%%%%%%%%%%%%%%%%%%%%%%%%%%%%%%%%%%%%%%%%%%%%%%%%%%%%%%%%%%%%%%%%%%%%%%%%%

\bigskip

    It is not difficult to make the simplification $\psi_{ft}(\vec w, \vec r~)=\psi_{ft}(\vec w)$
of Eq. (\ref{eq:psift})
more flux-tube-like. In the case where $\psi_{ft}$ depends on $\vec w$ and $\vec r$, 
Eqs. (\ref{eq:xi4})-(\ref{eq:xift}) become
\begin {eqnarray}
\xi^{QM}_{Q \bar q q \bar d}(w)&=&
{1 \over 2m_{Q}}\langle P_{Q \bar q q \bar d}({\vec P_{cm} \over 2}) \vert 
\bar Q  Q \vert  P_{Q \bar q q \bar d}(-{\vec P_{cm} \over 2}) \rangle \\
&=&
\int d^3w d^3r \psi^*_{Q \bar d}(\vec r~) \psi^*_{ft}(\vec w, \vec r~)
e^{-2i m_q \vec V_{cm} \cdot \vec w}
e^{-i m_d \vec V_{cm} \cdot \vec r}
\psi_{Q \bar d}(\vec r~) \psi_{ft}(\vec w, \vec r~)
\label{eq:xi4wr}
\end {eqnarray}
or, defining
\begin{equation}
\xi^{QM}_{Q \bar q q \bar d}(w)=1-\rho^2_{Q \bar q q \bar d}(w-1)
\end{equation}
we have
\begin{eqnarray}
\rho^2_{Q \bar q q \bar d} 
&=& {1 \over 3}
\int d^3w d^3r (2m_q \vec w + m_d \vec r~)^2\vert \psi_{Q \bar d}(\vec r~) \psi_{ft}(\vec w, \vec r~) \vert^2 \\
&=& {1 \over 3}\langle (2m_q \vec w + m_d \vec r~)^2 \rangle
\label{eq:rhoqbarq}
\end{eqnarray}
which reduces to the simplified results of the text in the appropriate limits.

   Now consider a generic example of a more realistic $\psi_{ft}$ that has a flux tube's shape:
\begin{equation}
\psi_{ft}(\vec w, \vec r~)={\beta_{ft} \over \pi^{1/2} }e^{-{1 \over 2}\beta^2_{ft} w^2_{\perp}}t(\vec w \cdot \hat r)
\end{equation}
where $\hat r=\vec r/r$, $\vec w_{\perp}=\vec w - (\vec w \cdot \hat r)\hat r$,
and $t(\vec w \cdot \hat r)$, which depends only on the longitudinal variable $\vec w \cdot \hat r$,
is a normalized tube-like function. ({\it E.g.}, one might have $t={1 \over \sqrt{r}} 
\theta (\vec w \cdot \hat r) \theta (r-\vec w \cdot \hat r)$ to create a cylindrical wavefunction
that is gaussian transverse to $\vec r$ and constant between $Q$ and $\bar d$.) With such a wavefunction
\begin{equation}
\rho^2_{Q \bar q q \bar d}={1 \over 3}[4m_q^2\langle w^2_{\perp}\rangle 
+ \langle (2m_q \vec w \cdot \hat r + m_d  r)^2 \rangle]
\label{eq:pluslongw} 
\end{equation}
The first term is as expected intuitively: it is the unchanged $\vec w_{\perp}$ part of the result of the text.
One might also naively interpret the second term as the old $\vec r$ term plus a new longitudinal
contribution due to the assumed spatial distribution of $\vec w \cdot \hat r$ in a tube-like configuration along
$\vec r$.

     I believe that the physics is more subtle than this. Consider the origin of
$\rho^2_{wf}$ in the nonrelativistic kinematics of our model. In the $Q \bar d$ sector
(see Eq. (\ref{eq:defR})), $ \vec r_Q = - m_d \vec r/ m_Q$ 
in the center of mass frame as $m_Q \rightarrow \infty$, while in $Q \bar q q \bar d$
(see Eq. (\ref{eq:defR'})), $\vec r_Q \rightarrow - (2m_q \vec w  + m_d \vec r~)/ m_Q $.
Since nonrelativistically
$\rho^2={1 \over 3}m_Q^2 \langle r_Q^2 \rangle$, we see that 
Eq. (\ref{eq:rhoQbard}) for $\rho^2_{Q  \bar d}$ and Eq. (\ref{eq:rhoqbarq}) for
$\rho^2_{Q \bar q q \bar d}$ are simply consequences of these nonrelativistic relations.

   To see the dangers of this approximation when the string tension and its renormalization
are large compared to $m_q$, consider the calculation of the mass of a system of heavy quarks
$Q$ and $\bar d$ at separation $\vec r$ connected by a renormalized flux tube, {\it i.e.}, 
of the state (\ref{eq:mixedstate}) of the text which has string tension
$b_{N_f}$ since it has the appropriate admixture
of $q \bar q$ pairs. If its mass were determined nonrelativistically one would obtain
\begin{equation}
M^{nr}-m_Q =m_d+{{2m_q c_{q \bar q}^2} \over {1+ c_{q \bar q}^2}}
\end{equation}
{\it i.e.},  the effective mass opposite $Q$ (against which it must recoil to conserve the position of the
center of mass)
would be  the probability-weighted masses 
of the pure $ \bar d$ state and the $ \bar q q \bar d$ admixture. On adding interactions (both
the diagonal potential $b_0r$ and the off-diagonal perturbation $H_{pc}^{q \bar q}$
which mixes $Q\bar d$ and $Q \bar q q \bar d$), we obtain
the correct answer 
\begin{equation}
M^{adiabatic}-m_Q =m_d+b_{N_f}r~~.
\end{equation}
Thus the mass $2m_q$ does not in this circumstance have an
independent reality as indicated by the nonrelativistic kinematics just described:
it is subsumed into the properly renormalized string tension. Indeed, it
is an implicit assumption of the model for the $q \bar q$ pairs described in
the text that the unquenched flux tube also behaves like a relativistic
string: it should support (only) transverse waves moving with the speed of light.
Thus while the $q \bar q$ pairs change the longitudinal distribution of
$\vec r_Q $, {\it this effect is already described in the flux tube
model by the mass $b_{N_f}r$ residing in the flux tube}:
when $r$ increases, not only does $\bar d$ move, but
so does the center of mass of the flux tube. It would therefore be double-counting to include
the effect on $\langle r^2_Q \rangle$ of the $2m_q\vec w \cdot \hat r$ term
of Eq. (\ref{eq:pluslongw}).
I should hasten to add that the nonrelativistic quark model used in this paper does not
{\it explicitly} take this effect into account. To do so would require a 
full treatment of the flux tube degrees of freedom ({\it versus} the adiabatic, potential-model
approximation used here). Nevertheless, even the nonrelativistic quark model 
has undoubtedly already taken some of this effect 
into account implicitly by its choice of such free parameters as $m_d$. 
({\it E.g.}, in many applications the constituent quark mass is effectively
$m_d=m^0_d+{1 \over 2}b \langle r \rangle$). In contrast, there is no mechanism
in the quark potential model to take into account transverse displacements
of $Q$ relative to $\vec r$ ~~\cite{NIonTransverse}. These transverse displacements
are the true degrees of freedom of the (quenched and unquenched) flux tubes and the
reaction of $Q$ to them makes a non-potential-model-type transverse contribution to
$\rho^2$.

    By renormalizing the string tension from $b_0 \rightarrow b_{N_f} < b_0$,
$q \bar q$ pairs 
increase the longitudinal contribution to $\rho^2$
(at least in the nonrelativistic approximation ${1 \over 2}b  r << m_d$). However, 
this increase is {\it not} compensated by
nonresonant production, since it 
is this same string tension $b_{N_f} < b_0$ which controls the structure
and thereby the production of excited resonances. {\it I.e.}, the longitudinal effect of the pairs is real,
but it simply renormalizes resonance physics. The dynamics behind this balancing
act, characteristic of the adiabatic approximation, can be seen by calculating 
the contribution of the flux tube to the Isgur-Wise function \cite{NIonTransverse}:
since the flux tube has only transverse internal degrees of freedom, it has no impact on
longitudinal overlap integrals
over the total separation $\vec r$.
Based on the preceding arguments, the transverse contributions to $\langle r^2_Q \rangle$
would also be those of a relativistic string with string tension $b_0$ or $b_{N_f}$ in the quenched and
unquenched flux tubes, respectively. In contrast to the longitudinal contribution to
$\langle r^2_Q \rangle$, the transverse contributions correspond to the reaction of $Q$
to real internal degrees of freedom, and these degrees of freedom (both gluonic and $q \bar q$)
can be excited by the action of the $\bar Q \Gamma Q$ current. When acting on the pure $Q \bar d$ piece
of the state, the current excites hybrid mesons which in the quenched limit exactly compensate for the loss of
rate from the elastic channel due to transverse contributions to  $\langle r^2_Q \rangle$~~\cite{NIonTransverse}.
In the $Q \bar q q \bar d$ sector, the current could in principle excite
either of the strings internal to mesons $(Q \bar q)_a$ or $(q \bar d)_b$, or it could
excite the center of mass of the $q \bar q$ pair. 
In the quark potential model approximation to this latter process, 
which is of course the one of interest for this paper, one
would recover the result shown in Eq. (\ref{eq:pluslongw}) less the longitudinal part of $\vec w$:
\begin{eqnarray}
\rho^2_{Q \bar q q \bar d}&=&{4m_q^2 \over 3}\langle w^2_{\perp}\rangle 
+ {m_d^2 \over 3}\langle   r^2 \rangle \\
&=&{4m_q^2 \over {3\beta^2_{ft}}} + {m_d^2 \over {3\beta^2_{Q\bar d}}}~.
\label{eq:transverserho}
\end{eqnarray}
The first term of this formula differs from the expression of the text by the factor 
${2 \over 3}$ corresponding
to two of the three degrees of freedom of $\vec w$ being active. Note that for this picture to be consistent,
the total transverse contribution to $\langle r^2_Q \rangle$ must be that of a relativistic string
with string tension $b_{N_f}$; the decomposition into $Q \bar d$ and $Q \bar q q \bar d$
components is only useful in identifying the compensating channels required by Bjorken's sum rule.
However, in the renormalized string picture $\langle r^2_Q \rangle$ of course depends
just on $b_{N_f}$, while Eq. (\ref{eq:transverserho}) shows a contribution proportional to $2m_q$.
It would be interesting to examine how the dynamics of pair creation in the flux tube leads to
such a term. I speculate that the mechanism is the ``consumption" of a piece of flux tube
of length $\Delta r \sim 2 m_q/b$ in a nonlocal pair creation process.

    In summary, in this Appendix I have described several subtleties in the description of 
$q \bar q$ pair creation in the flux tube model, and pointed out some interesting issues
which arise in the physics of the renormalized flux tube.
While I believe these matters are important conceptually and 
are worthy of further study, I am convinced that other uncertainties described in the text
are of far greater impact numerically on our results. Given this and the great
convenience of the spherical approximation, I therefore chose this simpler if less basic
framework for the discussion of the text.

\vfill\eject

%%%%%%%%%%%%%%%%%%%%%%%%%%%%%%%%%%%%%%%%%%%%%%%%%%%%%%%%%%%%%%%%%%%%%%%%%%
{\noindent \bf APPENDIX B: Determining $\eta_{q \bar q}$}
%%%%%%%%%%%%%%%%%%%%%%%%%%%%%%%%%%%%%%%%%%%%%%%%%%%%%%%%%%%%%%%%%%%%%%%%%%

\bigskip

    Eq. (\ref{eq:Hpc}) defines the action of the pair creation Hamiltonian on $\vert P_{Q \bar d} \rangle$.
This perturbation not only produces pairs to make the eigenstate $\vert P_Q \rangle$ of Eq. (\ref{eq:mixedstate}),
but also leads to strong decays. In particular, the projection of the state (\ref{eq:defP4})
onto the continuum
states $\vert \vec P_{cm};ab(\vec q~) \rangle$ determines the 
$P_Q \rightarrow (Q \bar q)_a (q \bar d)_b$ coupling constants.
By heavy quark symmetry \cite{IWspec}, the same dynamics determine the 
$P^*_Q \rightarrow (Q \bar q)_a (q \bar d)_b$ coupling constants,
where $P^*_Q$ is the vector partner of the pseudoscalar state $P_Q$. In this Appendix I
use these facts to
determine the strength parameter $\eta_{q \bar q}$ of Eq. (\ref{eq:Hpc}) by comparing to the decays
$P^*_Q \rightarrow P_Q \pi$.

    I begin with a practical matter. As $m_Q \rightarrow \infty$, the decays $P^*_Q \rightarrow P_Q \pi$
are forbidden since $P^*_Q$ and $P_Q$ become degenerate heavy quark spin partners. However,
$\Gamma(K^* \rightarrow K \pi)$ is known, and $\Gamma(D^* \rightarrow D \pi)$ can be deduced
if one assumes that the successful phenomenology of magnetic dipole decays can be extended
to $\Gamma(D^* \rightarrow D \gamma)$. (The  total width of the $D^*$ is so small
that only decay branching fractions and not decay widths are known.  If one takes the
$K^* \rightarrow K \pi$ and $K^* \rightarrow K \gamma$ decays and scales them appropriately
in $m_Q$ assuming that heavy quark scaling works all the way down to $m_s$,
the observed $D^* \rightarrow D \pi$ and
$D^* \rightarrow D \gamma$ branching ratios are explained nearly perfectly. This is
another example of the often-noted fact that in many circumstances a strange quark
behaves like  heavy quark.) Since the branching ratio for $D^{*0} \rightarrow D^0 \pi^0$
is well-determined experimentally, I will use the value $\Gamma(D^{*0} \rightarrow D^0 \pi^0)
\simeq 30$ keV deduced in this manner as input ``data".

    The calculations themselves are simple. If $g_0 D^{0\dagger}\partial_{\mu} \pi^0 D^{0* \mu}$
is the effective Lagrangian density for the decay leading to $\Gamma(D^{*0} \rightarrow D^0 \pi^0)=
{{g^2q^3} \over {48 \pi m^2_{D^*}}}$, then in the center of mass, with
the pion emitted with momentum $\vec q$ from a $D^*$ with polarization $+1$ along $\hat z$,
\begin{eqnarray}
-{{ig_0 q_+} \over {\sqrt{2} (2 \pi )^{9/2}}} &=& \eta_{q \bar q}\sqrt{1 \over 3}
{{\tilde m_D \tilde m_{\pi}^{1/2} \beta^3_D \beta^{3/2}_{\pi} \beta^{3/2}_{ft}} \over {8 \pi^6}}
\int d^3r \int d^3v \int d^3w \nonumber \\
&&
e^{-{1 \over 2} \beta^2_D(\vec w + {1 \over 2} \vec v)^2
-{1 \over 2} \beta^2_{\pi}(\vec r- \vec w + {1 \over 2} \vec v)^2
-{1 \over 2} \beta^2_{ft} w^2
-{1 \over 2} \beta^2_D r^2}
{1 \over {(2 \pi)^{3/2}}}e^{-i{\vec q \over 2} \cdot (\vec r + \vec w - {1 \over 2} \vec v)}
\psi_{pc}(\vec v)_+
\end{eqnarray}
in which the ${1 \over \sqrt{2}}$ for the pure $\pi^0$ decay via $\eta_{u \bar u}$ has been
explicitly included, $\tilde m_X$ is the ``mock meson" mass given by the sum of the
constituent quark masses, and 
$\psi_{pc}(\vec v)_+={\beta^{5/2}_{pc} \over \pi^{3/4}}v_+e^{-{1 \over 2} \beta^2_{pc} v^2}$.
The  integrals are straightforward and give
\begin{equation}
g_0=-\eta_{q \bar q}
{
{16 \pi^{3/4} \tilde m_D \tilde m_{\pi}^{1/2} 
\beta^3_D \beta^{3/2}_{\pi} \beta^{3/2}_{ft} \beta^{5/2}_{pc}({1 \over 2}-a_{DD} -b_{DD})} 
\over 
{\sqrt{3}{\beta_v}_{DD}^5 {\beta_y}_{DD}^3 {\beta_w}_{DD}^3 }
}
e^{-{q^2 \over {8\bar \beta_{DD}^2}}}
\end{equation}
where
\begin{eqnarray}
{1 \over {\bar \beta_{DD}^2}}&=&
{1 \over {\beta^2_y}}+
{1 \over {\beta^2_w}} \bigl( {{\beta^2_D+2 \beta^2_{\pi}} \over {\beta^2_D+\beta^2_{\pi}}} \bigr)^2+
{1 \over {\beta^2_v}}({1 \over 2}-a-b)^2\\
{\beta_v}_{DD}^5&=&
\beta^2_{pc}
+a^2(\beta^2_D+\beta^2_{\pi}+\beta^2_{ft})
+(b^2+{1 \over 4})(\beta^2_D+\beta^2_{\pi})
+a\beta^2_D+
(b-2ab-a)\beta^2_{\pi} \\
{\beta_y}_{DD}^2&=&  \beta^2_D+\beta^2_{\pi} 
\label{eq:betay}  \\
{\beta_w}_{DD}^2&=& \beta^2_D + \beta^2_{ft} +{{\beta^2_{\pi} \beta^2_D} \over {\beta^2_D+\beta^2_{\pi}}}    \\
a_{DD}&=&-
{\beta^4_D \over {2 
[(\beta^2_D+\beta^2_{\pi}) (\beta^2_D+\beta^2_{\pi}+\beta^2_{ft}) -  \beta^4_{\pi}]}}
\\
b_{DD}&=&-
{{\beta^2_{\pi}(2\beta^2_D+\beta^2_{ft}) } \over {2 
[(\beta^2_D+\beta^2_{\pi}) (\beta^2_D+\beta^2_{\pi}+\beta^2_{ft}) -  \beta^4_{\pi}]}}~~.
\label{eq:b}
\end{eqnarray}

    From the ``measured" $D^{0*} \rightarrow D^0 \pi^0$ width, we can deduce that $g_0 \simeq 11$.
With our canonical parameters it follows that, with $\beta_{pc}$ expressed in GeV,
\begin{equation}
\eta_{q \bar q} \simeq
0.32[{{3+2f} \over f^{1/2}}]^{3/2}[{{\beta_{pc}^2 + 0.06} \over \beta_{pc}}]^{5/2}
\end{equation}
where since $\beta_D \simeq \beta_{\pi} \simeq \beta_{ft}/f^{1/2}$ I have been able
to explicitly display the dependence on $f$ as well as $\beta_{pc}$. We see that for
a variation of $\pm 1$ around the canonical value $f=2$, $\eta_{q \bar q}$ varies
by less than $10 \%$. Thus we conclude that (since $\beta^2_{pc} >> 0.06$ GeV$^2$),
\begin{equation}
\eta_{q \bar q} \simeq 0.9
[{\beta_{pc} \over 0.58~ \rm{GeV}}]^{5/2}~ \rm{GeV}
\end{equation}
{\it i.e.}, $\eta_{q \bar q} \simeq 0.9$ for the canonical value $\beta_{pc}=0.58$ GeV. We also
see from this formula the expected result
that as the pair creation operator becomes more pointlike, $\eta_{q \bar q} \rightarrow \infty$.

\vfill\eject

%%%%%%%%%%%%%%%%%%%%%%%%%%%%%%%%%%%%%%%%%%%%%%%%%%%%%%%%%%%%%%%%%%%%%%%%%%
{\noindent \bf APPENDIX C: Calculating $\tau^{(m)}_{1/2}$ and $\tau^{(p)}_{3/2}$ 
for Selected Low-lying Exclusive Nonresonant Channels}
%%%%%%%%%%%%%%%%%%%%%%%%%%%%%%%%%%%%%%%%%%%%%%%%%%%%%%%%%%%%%%%%%%%%%%%%%%

\bigskip

    As discussed in Ref. \cite{IWonBj}, the semileptonic decays
$\bar B \rightarrow D^{(n)}_{{s'}_{\ell}^{{\pi '}_{\ell}}} \ell \bar \nu_{\ell}$
are governed in the heavy quark limit by generalized Isgur-Wise functions
which determine all of the form factors for the decay of the $\bar B$
with $s_{\ell}^{\pi_{\ell}}={1\over 2}^-$ to both of the states of a heavy quark spin multiplet with
quantum numbers ${s'}_{\ell}^{{\pi '}_{\ell}}$. As described in the text and elsewhere \cite{IWonBj},
\begin{equation}
\rho^2={1\over 4}+\Delta \rho^2_{pert}+\rho^2_{Q\bar d}+\Delta \rho^2_{sea}
\end{equation}
where the ${1\over 4}$ is Bjorken's relativistic correction \cite{Bj},
$\Delta \rho^2_{pert}$ is a perturbative QCD radiative correction, and
$\rho^2_{Q\bar d}$ and $\Delta \rho^2_{sea}$ are the contributions to the slope of
$\xi(w)$ from the valence and sea quarks, respectively. As we have seen, these latter
two contributions may be related to the rates of decay into inelastic channels by
\begin{eqnarray}
\rho^2_{Q\bar d}&=& 
\sum_{m={1\over 2}^+ Q \bar d~resonances}\vert \tau^{(m)}_{1/2} \vert^2
+2 \sum_{p={3\over 2}^+ Q \bar d~resonances}\vert \tau^{(p)}_{3/2} \vert^2  \\
\rho^2_{sea}&=&
\sum_{m={1\over 2}^+ continuum}\vert \tau^{(m)}_{1/2} \vert^2
+2 \sum_{p={3\over 2}^+ continuum}\vert \tau^{(p)}_{3/2} \vert^2
\label{eq:tauplustau}  
\end{eqnarray}
where the $\tau$'s are the appropriate Isgur-Wise functions. 
Once the $\tau$'s are specified, all transition form factors to the states of
a heavy quark spin multiplet may be determined from symmetry considerations.
For this reason, it is useful to calculate the $\tau$'s in the simplest possible
setting, namely for the case where $b$ and $c$ are {\it spinless}. This is
possible since the $\tau$'s depend only on the dynamics of the light
degrees of freedom, {\it i.e.}, on the transition 
$s_{\ell}^{\pi_{\ell}}={1\over 2}^- \rightarrow {s'}_{\ell}^{{\pi '}_{\ell}}={1\over 2}^+$ or ${3\over 2}^+$.

    Another key simplification arises from the dynamics of the pair creation
process. As demonstrated in the text, when
$b(\vec v) \rightarrow c(\vec {v'})$ in a $Q \bar q q \bar d$ state, excitation of the variable $\vec r$
cannot lead to a contribution of order $(w-1)$. Furthermore,
the variable $\vec v$ cannot be excited since this is a $q \bar q$ internal coordinate. 
Thus to contribute at order $(w-1)$, $b(\vec v) \rightarrow c(\vec {v'})$ {\it must} 
kick the $\vec w$ coordinate into an $\ell_w=1$ state: {\it  such a state
is the parent of all nonresonant production to order} $(w-1)$. Thus
$\Delta \rho^2_{sea}$ arises entirely from the ``decay" of the lowest $\ell_w=1$ excited 
state of $c \bar q q \bar d$ arising from the $b \rightarrow c$ transition from $b \bar q q \bar d$.
With the $q \bar q$ pair in $J^P=0^+$, the decay can thus occur from the six states
$w_+ \uparrow$, 
$w_+ \downarrow$, 
$w_0 \uparrow$, 
$w_0 \downarrow$, 
$w_- \uparrow$, and
$w_- \downarrow$,
depending on which component of $\vec w$ is excited by the recoil. Here
$w_+, w_0, w_-$ represent the components of  the $\ell_w=1$ state, and  
$\uparrow, \downarrow$ represent the spin state of the $\bar d$ spectator quark.
Since the total decay rate of $b \bar q q \bar d \uparrow$ and $b \bar q q \bar d \downarrow$
must be the same, we can simplify if we average rates over the initial $\bar d$ spin
and over directions of $\vec P_{cm}$ (or equivalently, over directions of $\vec w$).
Then since the average over the six states just listed will be the same as the average over
the two $j={1\over 2}^+$ and four $j={3\over 2}^+$ states formed from them, we can
deal with ``parent" states that are states with good angular momenta and which therefore
uniquely feed the ${s'}_{\ell}^{{\pi '}_{\ell}}={1\over 2}^+$ and ${3\over 2}^+$ states,
respectively.

    Let me provide an example: the production of the $(D+D^*)\pi$
nonresonant states. Since we know that all of $\Delta \rho^2_{sea}$ arises from
the ``parent" state, the fraction of $\Delta \rho^2_{sea}$ coming from a given channel
can just be obtained as the $jm_j$ average of the square of an overlap between a given $jm_j$ in
$Q \bar q q \bar d$ and the two particle continuum state of interest. Thus from the
$jm_j={1\over 2}{1\over 2}$ state we can extract 
$\langle (D+D^*)_{{1 \over 2}^- {1 \over 2}} \pi (\vec q~) \vert c \bar q q \bar d ;{1\over 2}^+{1\over 2} \rangle$,
which ought to leave the $(D+D^*)\pi$ system in an $S$-wave. Explicitly, 
as $m_Q \rightarrow \infty$,
\begin{eqnarray}
\langle (D+D^*)_{{1 \over 2}^- {1 \over 2}} \pi (\vec q~) \vert c \bar q q \bar d ;{1\over 2}^+{1\over 2} \rangle
&=&
\int d^3w \int d^3v \int d^3r
{\beta_D^{3/2} \over \pi^{3/4}}e^{-{1 \over 2}\beta_D^2(\vec w+ {1 \over 2}\vec v)^2} 
{\beta_{\pi}^{3/2} \over \pi^{3/4}}e^{-{1 \over 2}\beta_{\pi}^2(\vec r-\vec w+ {1 \over 2}\vec v)^2}
  \nonumber \\
&& 
{1 \over {(2\pi)^{3/2}}}e^{-i{\vec q \over 2} \cdot (\vec r+ \vec w - {1 \over 2}\vec v)} 
{\tilde \beta_{pc}^{5/2} \over \pi^{3/4}}e^{-{1 \over 2}\beta_{pc}^2 v^2} 
{\beta_{ft}^{5/2} \over \pi^{3/4}}e^{-{1 \over 2}\beta_{ft}^2 w^2} 
{\beta_B^{3/2} \over \pi^{3/4}}e^{-{1 \over 2}\beta_B^2 r^2}
  \nonumber \\
&& 
  \nonumber \\
&& 
 \Sigma (\vec w, \vec v)
\end{eqnarray}
where
\begin{equation}
\Sigma \equiv \langle   \uparrow \sqrt{1 \over 2} (\uparrow \downarrow - \downarrow \uparrow)  \vert
    \sqrt{1 \over 3} 
(\uparrow \uparrow v_--[\uparrow \downarrow + \downarrow \uparrow]v_z- \downarrow \downarrow v_+)
\sqrt{2 \over 3}(-w_+\downarrow-w_z\uparrow)  \rangle
\label{eq:Dpiintegral}
\end{equation}
is the spin overlap matrix element of the three quarks $\bar q q \bar d$, respectively. We get
\begin{equation}
\langle (D+D^*)_{{1 \over 2}^- {1 \over 2}} \pi (\vec q~) \vert c \bar q q \bar d ;{1\over 2}^+{1\over 2} \rangle
=-I^{00}_{+-} - I^{00}_{zz}
\end{equation}
where
\begin{equation}
I^{00}_{ij} \equiv {1 \over 3}
{
{\beta_D^{3/2}\beta_{\pi}^{3/2}\beta_{ft}^{5/2}\tilde \beta_{pc}^{5/2}\beta_B^{3/2}}
\over
{\pi^{15/4}}
}
\int d^3w \int d^3v \int d^3r
{{v_iw_j} \over {(2\pi)^{3/2}}}e^{-i{\vec q \over 2} \cdot (\vec r+ \vec w - {1 \over 2}\vec v)-{1 \over 2}E^2} 
\end{equation}
with
\begin{equation}
E^2=
\beta_D^2 (\vec w + {1 \over 2}\vec v)^2
+\beta_{\pi}^2 (\vec r - \vec w + {1 \over 2}\vec v)^2
+\beta_{ft}^2 w^2
+\tilde \beta_{pc}^2 v^2
+\beta_B^2 r^2~.
\end{equation}
In these formulas I have distinguished between $\beta_D^2$ and $\beta_B^2$ to allow the ``ancestry"
of terms to be traced, even though $\beta_B=\beta_D$ from heavy
quark symmetry. This integral is easily done, giving
\begin{equation}
I^{00}_{ij}=I^{00} \Bigl[ c^{00} \delta_{ij} + c^{00}_{ij} {{q_iq_j} \over \beta^2_v}  \Bigr]
e^{-q^2/ 8\tilde{\bar \beta}^2}
\end{equation}
where $I^{00}$, $c^{00}_{ij}$, $c^{00}$, and $\tilde{\bar \beta}$ are given below. For the problem at hand, we get
\begin{eqnarray}
\langle (D+D^*)_{{1 \over 2}^- {1 \over 2}} \pi (\vec q~) \vert c \bar q q \bar d ;{1\over 2}^+{1\over 2} \rangle
&=&-\Bigl[3c^{00}+ c^{00}_{ij}{{q_+q_-} \over \beta^2_v} +c^{00}_{ij}{q^2_z \over \beta^2_v}  \Bigr] 
I^{00}e^{-q^2/ 8\tilde{\bar \beta}^2}\nonumber \\
&=&-\Bigl[3c^{00} + c^{00}_{ij} {{q^2} \over \beta^2_v}   \Bigr] 
I^{00} e^{-q^2/ 8\tilde{\bar \beta}^2} \nonumber \\
&=&-\sqrt{4 \pi} \Bigl[3c^{00} + c^{00}_{ij} {{q^2} \over \beta^2_v}  \Bigr] 
I^{00} e^{-q^2/ 8\tilde{\bar \beta}^2}Y_{00}(\Omega_q)~, 
\end{eqnarray}
a pure $S$-wave decay as required, with partial wave amplitude 
$A_{1 / 2} \equiv -\sqrt{4 \pi} \Bigl[3c^{00} + c^{00}_{ij} {{q^2} \over \beta^2_v}  \Bigr] I^{00}
e^{-q^2/ 8\tilde{\bar \beta}^2}$.
Note that with $Y_{00}$ factored out, $\vert A_{1 / 2} \vert^2$ is already the probability
for this channel integrated over angles $\Omega_q$, leaving only an integral 
$\int dq q^2 \vert A_{1 / 2} \vert^2$ to be done to sum over all 
$(D+D^*) \pi $ states with quantum numbers ${1 \over 2}^+ {1 \over 2}$ at any fixed value of $(w-1)$.

    Next consider 
$\langle (D+D^*)_{{1 \over 2}^- {1 \over 2}} \pi (\vec q~) \vert c \bar q q \bar d ;{3\over 2}^+{3\over 2} \rangle$,
which proceeds by replacing $\sqrt{2 \over 3}(-w_+\downarrow-w_z\uparrow)  \rangle$ by $-w_+ \uparrow$
in Eq. (\ref{eq:Dpiintegral}). With this change one gets
\begin{eqnarray}
\langle (D+D^*)_{{1 \over 2}^- {1 \over 2}} \pi (\vec q~) \vert c \bar q q \bar d ;{3\over 2}^+{3\over 2} \rangle
&=&-\sqrt{3 \over 2} I^{00}_{z+} \nonumber \\
&=&-\sqrt{3 \over 2} c^{00}_{ij} {q_zq_+ \over \beta^2_v}I^{00} e^{-q^2/ 8\tilde{\bar \beta}^2}          \nonumber \\
&=&-\sqrt{1 \over 5}[\sqrt{4 \pi} c^{00}_{ij} {{q^2} \over \beta^2_v} 
I^{00}e^{-q^2/ 8\tilde{\bar \beta}^2}] Y_{21}(\Omega_q)~, 
\end{eqnarray}
a pure $D$-wave decay as required. Moreover, since $-\sqrt{1 \over 5}$ is the Clebsch-Gordan coefficient
for coupling  $(D+D^*)_{{1 \over 2}^- {1 \over 2}}$ and an  $\ell=2$, $m=1$ pion into a ${3 \over 2}{3 \over 2}$
state, we can deduce that decays to this whole angular momentum multiplet 
with $(D+D^*) \pi $ in a $D$-wave coupled to ${s'_{\ell}}^{{\pi '}_{\ell}}={3 \over 2}^+$
are controlled by a $D$-wave amplitude 
$A_{3 / 2} \equiv -\sqrt{4 \pi} c^{00}_{ij} {{q^2} \over \beta^2_v} I^{00}e^{-q^2/ 8\tilde{\bar \beta}^2}$.

   To complete this pedagogical example, I note that since the 
partial wave decay amplitudes are independent
of the magnetic substate $m$,
\begin{eqnarray}
{1 \over 6} \sum_m \vert A_{{1 / 2}}   \vert^2 &=& {1 \over 3}\vert A_{{1 / 2}}   \vert^2
\label{eq:A1/2rate} \\
{1 \over 6} \sum_m \vert A_{{3 / 2}}   \vert^2 &=& {2 \over 3}\vert A_{{3 / 2}}   \vert^2 ~~,
\label{eq:A3/2rate}
\end{eqnarray}
where the ${1 \over 6}$ arises from averaging over the six $jm_j$ states.
On comparison with Eq. (\ref{eq:tauplustau}), we see that 
$\tau_{1 / 2}=\sqrt{1 \over 3} A_{{1 / 2}} \sqrt{{1 \over 3} \Delta \rho^2_{sea} }$ and
$\tau_{3 / 2}=\sqrt{1 \over 3} A_{{3 / 2}} \sqrt{ {1 \over 3} \Delta \rho^2_{sea} }$. Note that 
${1 \over 3} \Delta \rho^2_{sea}$ appears since the 
overlap amplitudes $A_{1 / 2}$ and $A_{3 / 2}$ as calculated are for a single flavor,
while the factor $\sqrt{1 \over 3}$ arises as a residue of the angular and spin averaging.

    The tricks outlined here are more powerful for more complex decays. I will give one
partial illustration:
$\langle [(D+D^*) \rho]_{{3\over 2}^+{3\over 2}} 
\vert c \bar q q \bar d ;{3\over 2}^+{3\over 2} \rangle$, where the subscripts
on the bracket are total spin quantum numbers, but do not include relative
orbital angular momentum. The overlap integral for this decay
is obtained by replacing the pion spin wavefunction 
$\sqrt{1 \over 2}(\uparrow \downarrow- \downarrow \uparrow)$ by $\uparrow \uparrow$,
$\sqrt{2 \over 3}(-w_+ \downarrow-w_z \uparrow)$ by $-w_+ \uparrow$,  and $\beta_{\pi}$
by $\beta_{\rho}$ in Eq. (\ref{eq:Dpiintegral}) to give
\begin{eqnarray}
\langle [(D+D^*) \rho]_{{3\over 2}^+{3\over 2}} 
 \vert c \bar q q \bar d ;{3\over 2}^+{3\over 2} \rangle
&=&-\sqrt{3} I^{00}_{+-} \nonumber \\
&=&-\sqrt{3} [ 2 c^{00} + c^{00}_{ij} {q_+q_- \over \beta^2_v}]I^{00} e^{-q^2/ 8\tilde{\bar \beta}^2}  \nonumber \\
&=&-\sqrt{4 \over 3}\sqrt{4 \pi}
\Bigl([ 3 c^{00} + c^{00}_{ij} {q^2 \over \beta^2_v} ] Y_{00}   \nonumber \\
&&~~~~~~~- c^{00}_{ij} {q^2 \over \beta^2_v} \sqrt{1 \over 5} Y_{20} \Bigr)I^{00}e^{-q^2/ 8\tilde{\bar \beta}^2} ~, 
\end{eqnarray}
By examining the Clebsch-Gordan coefficients for coupling the spin state
$[(D+D^*) \rho]_{{3\over 2}^+{3\over 2}}$ to a relative orbital   $S$-wave 
or a $D$-wave to get  a ${3 \over 2}{3 \over 2}$
state, we can deduce that the $S$- and $D$-wave amplitudes for this decay are
$- \sqrt{4 \over 3}[\sqrt{4 \pi} (3 c^{00} + c^{00}_{ij} {q^2 \over \beta^2_v} )I^{00}e^{-q^2/ 8\tilde{\bar \beta}^2}]$ and
$\sqrt{4 \over 3}[\sqrt{4 \pi} c^{00}_{ij} {q^2 \over \beta^2_v} I^{00}e^{-q^2/ 8\tilde{\bar \beta}^2}]$,
respectively. One very simple overlap
integral thus gives two partial wave amplitudes simultaneously.

    A complete set of results  are given in the text in Table II in terms of the
following basic integrals:
\begin{eqnarray}
I^{00}_{ij} &=& I^{00} 
\Bigl[c^{00} \delta_{ij} + c^{00}_{ij} {{q_iq_j} \over \tilde \beta^2_v}  \Bigr] e^{-{q^2 / {8 \tilde {\bar \beta}^2}}}   \\
I^{10}_{ijk} &=& I^{10} 
\Bigl[c^{10}_i q_i \delta_{jk} + 
c^{10}_j q_j \delta_{ik} +
c^{10}_k q_k \delta_{ij} +
c^{10}_{ijk} {{q_i q_j q_k} \over {\tilde \beta^2_{v~**}}} \Bigr] e^{-{q^2 / {8 \tilde {\bar \beta}_{**}^2}}}  \\
I^{01}_{ijk} &=& I^{01} 
\Bigl[c^{01}_i q_i \delta_{jk} + 
c^{01}_j q_j \delta_{ik} +
c^{01}_k q_k \delta_{ij} +
c^{01}_{ijk} {{q_i q_j q_k} \over {\tilde \beta^2_{v~a}}} \Bigr] e^{-{q^2 / {8 \tilde {\bar \beta}_a^2}}}   
\end{eqnarray}
where $I^{00}_{ij}$ is specified in terms of
\begin{eqnarray}
I^{00}&=&
{
{2 \beta^{3/2}_D \beta^{3/2}_B \beta^{3/2}_{\pi} \beta^{5/2}_{ft} \tilde \beta^{5/2}_{pc}} 
\over 
{3 \pi^{3/4}  \tilde \beta^5_v \beta^5_w \beta^3_y}
}   \\
c^{00}&=& +4a\beta^2_w    \\
c^{00}_{ij}&=&\Bigl[({1 \over 2}-a -b)
\bigl( {{\beta^2_B+2 \beta^2_{\pi}} \over {\beta^2_B+\beta^2_{\pi}}} \bigr)
-{a \beta^2_w \over \tilde \beta^2_v}({1 \over 2}-a -b)^2\Bigr] \\
\tilde \beta^2_v&=&
\tilde \beta^2_{pc}
+a^2(\beta^2_D+\beta^2_{\pi}+\beta^2_{ft})
+b^2(\beta^2_B+\beta^2_{\pi})               \nonumber  \\
&&+{1 \over 4}(\beta^2_D+\beta^2_{\pi})
+a\beta^2_D+
(b-2ab-a)\beta^2_{\pi}       \\
\beta_y^2 &=&  \beta^2_B+\beta^2_{\pi}   \\
\beta_w^2 &=& \beta^2_D + \beta^2_{ft} +{{ \beta^2_B \beta^2_{\pi} } \over {\beta^2_B+\beta^2_{\pi}}}    \\
a &=&-
{{(\beta^2_B+\beta^2_{\pi})(\beta^2_D-\beta^2_{\pi})+\pi^4} \over {2 
[(\beta^2_B+\beta^2_{\pi}) (\beta^2_D+\beta^2_{\pi}+\beta^2_{ft}) -  \beta^4_{\pi}]}}
\\
b&=&-
{{\beta^2_{\pi}(2\beta^2_D+\beta^2_{ft}) } \over {2 
[(\beta^2_B+\beta^2_{\pi}) (\beta^2_D+\beta^2_{\pi}+\beta^2_{ft}) -  \beta^4_{\pi}]}} \\
{1 \over { \tilde {\bar \beta}^2}}&=&
{1 \over {\beta^2_y}}+
{1 \over {\beta^2_w}} \bigl( {{\beta^2_B+2 \beta^2_{\pi}} \over {\beta^2_B+\beta^2_{\pi}}} \bigr)^2+
{1 \over {\tilde  \beta^2_v}}({1 \over 2}-a-b)^2~~.
\end{eqnarray}

    Similarly, $I^{10}_{ijk}$ is specified in terms of
\begin{eqnarray}
I^{10}&=& -
{
{4i \beta^{5/2}_{D^{**}} \beta^{3/2}_B \beta^{3/2}_{\pi} \beta^{5/2}_{ft} \tilde \beta^{5/2}_{pc}} 
\over 
{3 \pi^{3/4}  \tilde {\beta_v}_{**}^5 {\beta_w}_{**}^5 {\beta_y}_{**}^3}
}   \\
c^{10}_{i}&=&\Bigl[+a_{**}\bigl( {{\beta^2_B+2 \beta^2_{\pi}} \over {\beta^2_B+\beta^2_{\pi}}} \bigr)
-a_{**} ({1 \over 2}+a_{**}) ({1 \over 2}-a_{**} -b_{**}) 
{{\beta_w}_{**}^2 \over \tilde {\beta_v}_{**}^2} \Bigr] \\
c^{10}_{j}&=&\Bigl[-({1 \over 2}-a_{**} -b_{**})
-a_{**} ({1 \over 2}+a_{**}) ({1 \over 2}-a_{**} -b_{**}) 
{{\beta_w}_{**}^2 \over \tilde {\beta_v}_{**}^2} \Bigr] \\
c^{10}_{k}&=&\Bigl[+({1 \over 2}+a_{**})
\bigl( {{\beta^2_B+2 \beta^2_{\pi}} \over {\beta^2_B+\beta^2_{\pi}}} \bigr)
-a_{**} ({1 \over 2}+a_{**}) ({1 \over 2}-a_{**} -b_{**}) 
{{\beta_w}_{**}^2 \over \tilde {\beta_v}_{**}^2} \Bigr] \\
c^{10}_{ijk}&=&+{1 \over 4} \Bigl[({1 \over 2}-a_{**} -b_{**}) 
{\tilde {\beta_v}_{**}^2 \over {\beta_w}_{**}^2}
\bigl( {{\beta^2_B+2 \beta^2_{\pi}} \over {\beta^2_B+\beta^2_{\pi}}} \bigr)^2    \nonumber \\
&&
-({1 \over 2}+2a_{**})({1 \over 2}-a_{**} -b_{**})^2
\bigl( {{\beta^2_B+2 \beta^2_{\pi}} \over {\beta^2_B+\beta^2_{\pi}}} \bigr)         \nonumber \\
&&+a_{**} ({1 \over 2}+a_{**}) ({1 \over 2}-a_{**} -b_{**})^3 
{{\beta_w}_{**}^2 \over \tilde {\beta_v}_{**}^2} \Bigr]  
\end{eqnarray}
where 
$\tilde {\beta_v}_{**}^2$, ${\beta_y}_{**}^2$, ${\beta_w}_{**}^2$, 
$a_{**}$,  $b_{**}$, and  ${1 \over { \tilde {\bar \beta}_{**}^2}}$
are given by the formulas for the $I^{00}_{ij}$ variables
$\tilde \beta^2_v$, $\beta^2_y$, $\beta^2_w$, $a$,  $b$, and  ${1 \over { \tilde {\bar \beta}^2}}$, 
respectively, with $\beta_D \rightarrow \beta_{D^{**}}$ everywhere.

     Finally, $I^{01}_{ijk}$ is specified in terms of
\begin{eqnarray}
I^{01}&=& - 
{
{4i \beta^{3/2}_D \beta^{3/2}_B \beta^{5/2}_a \beta^{5/2}_{ft} \tilde \beta^{5/2}_{pc}} 
\over 
{3 \pi^{3/4} \tilde {\beta_v}_a^5 {\beta_w}_a^5 {\beta_y}_a^3}
}   \\
c^{01}_{i}&=&\Bigl[+ {a_a {\beta_w}_a^2 \over \tilde {\beta_v}_a^2} 
-{{a_a \beta^2_B} \over {{\beta^2_B+\beta^2_a}}} 
\bigl({{\beta^2_B+2 \beta^2_a} \over {\beta^2_B+\beta^2_a}} \bigr)
-a_a ({1 \over 2}-a_a+b_a) ({1 \over 2}-a_a -b_a) {{\beta_w}_a^2 \over \tilde {\beta_v}_a^2} \Bigr] \\
c^{01}_{j}&=&\Bigl[+{{\beta^2_B} \over {{\beta^2_B+\beta^2_a}}} ({1 \over 2}-a_a -b_a)
-a_a ({1 \over 2}-a_a+b_a) ({1 \over 2}-a_a -b_a) {{\beta_w}_a^2 \over \tilde {\beta_v}_a^2} \Bigr] \\
c^{01}_{k}&=&\Bigl[+\bigl({{\beta^2_B+2 \beta^2_a} \over {\beta^2_B+\beta^2_a}} \bigr)({1 \over 2}-a_a+b_a)
-a_a ({1 \over 2}-a_a+b_a) ({1 \over 2}-a_a -b_a) {{\beta_w}_a^2 \over \tilde {\beta_v}_a^2} \Bigr] \\
c^{01}_{ijk}&=&+{1 \over 4} \Bigl[
+({1 \over 2}-a_a -b_a) 
{\tilde {\beta_v}_a^2 \over {\beta_y}_a^2}
\bigl( {{\beta^2_B+2 \beta^2_a} \over {\beta^2_B+\beta^2_a}} \bigr)    
-{a_a {\beta_w}_a^2 \over {\beta_y}_a^2}({1 \over 2}-a_a -b_a)^2       \nonumber \\
&&
-{\tilde {\beta_v}_a^2 \over {\beta_w}_a^2}
\bigl( {{\beta^2_B+2 \beta^2_a} \over {\beta^2_B+\beta^2_a}} \bigr)^2
{{\beta^2_B} \over {{\beta^2_B+\beta^2_a}}} ({1 \over 2}-a_a -b_a)    \nonumber \\
&&
+{{ a_a \beta^2_B} \over {{\beta^2_B+\beta^2_a}}} 
\bigl( {{\beta^2_B+2 \beta^2_a} \over {\beta^2_B+\beta^2_a}} \bigr) ({1 \over 2}-a_a-b_a)^2   \nonumber \\
&&
-\bigl( {{\beta^2_B+2 \beta^2_a} \over {\beta^2_B+\beta^2_a}} \bigr)
({1 \over 2}-a_a+b_a)({1 \over 2}-a_a -b_a)^2   \nonumber \\
&&
+{a_a {\beta_w}_a^2 \over \tilde {\beta_v}_a^2}({1 \over 2}-a_a+b_a)({1 \over 2}-a_a -b_a)^3
\Bigr]  
\end{eqnarray}
where 
$\tilde {\beta_v}_a^2$, ${\beta_y}_a^2$, ${\beta_w}_a^2$, 
$a_a$,  $b_a$, and  ${1 \over { \tilde {\bar \beta}_a^2}}$
are given by the formulas for the $I^{00}_{ij}$ variables
$\tilde \beta^2_v$, $\beta^2_y$, $\beta^2_w$, $a$,  $b$, and  ${1 \over { \tilde {\bar \beta}^2}}$, 
respectively, with $\beta_{\pi} \rightarrow \beta_a$ everywhere.

\vfill\eject

{

\vfill\eject

\vspace{0.5cm}
\begin{table}[th]
  \caption[x]{$\tau^{(m)}_{1/2}$ and $\tau^{(p)}_{3/2}$ 
for nonresonant semileptonic decays to low-lying exclusive channels
in units of 
$\alpha_S   \equiv \sqrt{4 \pi} (3c^{00}+c^{00}_{ij}q^2/ \tilde \beta^2_v)
    I^{00}e^{-q^2/8 \tilde {\bar \beta}^2_v}$,
$\alpha_D   \equiv \sqrt{4 \pi} (c^{00}_{ij}q^2/ \tilde \beta^2_v)
    I^{00}e^{-q^2/8 \tilde {\bar \beta}^2_v}$,
$\beta^{10}_n   \equiv \sqrt{4 \pi} (c^{10}_{n}q)
    I^{10}e^{-q^2/8 \tilde {\bar \beta}^2_{**}}$,
$\gamma^{10}_n   \equiv \sqrt{4 \pi} (c^{10}_{ijk}q^3/ \tilde \beta^2_{v,**})
    I^{10}e^{-q^2/8 \tilde {\bar \beta}^2_{**}}$
$\beta^{01}_n   \equiv \sqrt{4 \pi} (c^{01}_{n}q)
    I^{01}e^{-q^2/8 \tilde {\bar \beta}^2_{a}}$, and
$\gamma^{01}_n   \equiv \sqrt{4 \pi} (c^{01}_{ijk}q^3/ \tilde \beta^2_{v,a})
    I^{01}e^{-q^2/8 \tilde {\bar \beta}^2_{a}}$,
as defined in Appendix C. Shown explicitly are the amplitudes for emission of a
$(D+D^*)$, $D^{**}_{3/2}$, or $D^{**}_{1/2}$ and a {\it positively charged light
hadron} ($\pi^+$, $\rho^+$, $a_2^+$, $a_1^+$, $b_1^+$, or $a_0^+$) 
from $\bar B^0$ by $\bar u u$ pair creation. 
(Here $(D+D^*)$, $D^{**}_{3/2}$, and $D^{**}_{1/2}$ are the lowest-lying
${s'_{\ell}}^{\pi'_{\ell}}={1 \over 2}^-$, ${3 \over 2}^+$, and ${1 \over 2}^+$ heavy quark spin
multiplets, respectively.)
The subscripts on a channel in the first column
are the total spin (the $s_a^{\pi_a}$ of the charmed meson $a$ plus the spin of the light
meson $b$), and the $ab$ relative orbital angular momentum, respectively; the second and third columns define the total
${s'_{\ell}}^{\pi'_{\ell}}$ as either ${1 \over 2}^+$ or ${3 \over 2}^+$.
Note that since partial wave amplitudes with respect to the direction of the vector $\vec q$
are given, full rates to a channel are obtained by integrating over $q^2dq$ and not $d^3q$,
{\it i.e.,} a factor $\sqrt{4 \pi}$ is included in each amplitude. Shown in
parentheses under each allowed amplitude is the fraction of the $\bar u u$ rate 
going into this channel in percent, based on Eqs. (\ref{eq:A1/2rate}) and (\ref{eq:A3/2rate}).}
  \begin{tabular}{lccc}
channel
 & $\sqrt{3} \tau_{1/2}^{(channel)} /\sqrt{\Delta \rho^2_{sea}/3}$  
  & $\sqrt{3} \tau_{3/2}^{(channel)} /\sqrt{\Delta \rho^2_{sea}/3}$ \\
 & (fraction of $\bar u u$ rate ($\%$))                               
  & (fraction of $\bar u u$ rate ($\%$))    \\
\hline
$[(D+D^*) \pi^+]_{{1 \over 2}S}$
 &    -$\alpha_S$    
  &         -       \\     
 & (0.4) 
  &     \\
$[(D+D^*) \pi^+]_{{1 \over 2}D}$       
 &          -  
  & $-\alpha_D$    \\              
  &                     
   & (1.5)    \\
$[(D+D^*) \rho^+]_{{3 \over 2}S}$      
 &          -                  
  & $-\sqrt{4 \over 3}\alpha_S$   \\
  &     
   & (1.0) \\
$[(D+D^*) \rho^+]_{{3 \over 2}D}$      
 & +$\sqrt{8 \over 3}\alpha_D$ 
  & $+\sqrt{4 \over 3}\alpha_D$  \\
 & (2.0) 
  & (2.0) \\
$[(D+D^*) \rho^+]_{{1 \over 2}S}$      
 & +$\sqrt{1 \over 3}\alpha_S$ 
  &         -    \\
 & (0.1) 
  &     \\
$[(D+D^*) \rho^+]_{{1 \over 2}D}$      
 &          -                  
  & $+\sqrt{1 \over 3}\alpha_D$   \\
 &     
  & (0.5) \\
$[D^{**}_{3/2} \pi^+]_{{3 \over 2}P}$  
 & $+ \sqrt{4 \over 3} [2\beta^{10}_j + 3\beta^{10}_i] + \sqrt{4 \over 3} \gamma^{10}$ 
  & $-\sqrt{10 \over 3} \beta^{10}_j - \sqrt{2 \over 15} \gamma^{10}$   \\
 & (0.5) 
  &    (0.8) \\
$[D^{**}_{3/2} \pi^+]_{{3 \over 2}F}$  
 & -
  & $+ \sqrt{6 \over 5} \gamma^{10}$   \\
 &  
  &    (0.1) \\
$[D^{**}_{1/2} \pi^+]_{{1 \over 2}P}$  
 & $+ \sqrt{2 \over 3} [ \beta^{10}_j - 3\beta^{10}_i - 2\beta^{10}_k] - \sqrt{2 \over 3} \gamma^{10}$ 
  & $-\sqrt{2 \over 3} [2\beta^{10}_j+3\beta^{10}_k] - \sqrt{2 \over 3} \gamma^{10}$ \\
 & (1.1) 
  &    (0.4) \\
$[D^{**}_{3/2} \rho^+]_{{5 \over 2}P}$  
 & -
  & $+\sqrt{12} [\beta^{10}_j+\beta^{10}_i] + \sqrt{48 \over 25} \gamma^{10}$  \\
 & 
  &    (2.6) \\
$[D^{**}_{3/2} \rho^+]_{{5 \over 2}F}$  
 & $ - \sqrt{16 \over 5} \gamma^{10}$
  & $- \sqrt{32 \over 25} \gamma^{10}$      \\
 & (0.1)
  &    (0.1) \\
$[D^{**}_{3/2} \rho^+]_{{3 \over 2}P}$  
 & $+ \sqrt{20 \over 9} [ 2\beta^{10}_j +  \beta^{10}_i ] + \sqrt{4 \over 5} \gamma^{10}$ 
  & $+\sqrt{2 \over 9} [ \beta^{10}_j-\beta^{10}_i] - \sqrt{2 \over 25} \gamma^{10}$ \\
 & (1.0) 
  &    (0.1) \\
$[D^{**}_{3/2} \rho^+]_{{3 \over 2}F}$  
 & -
  & $+ \sqrt{18 \over 25} \gamma^{10}$   \\  
 & 
  &    (0.1) \\
$[D^{**}_{3/2} \rho^+]_{{1 \over 2}P}$  
 & $+ \sqrt{16 \over 9} [  \beta^{10}_j -  \beta^{10}_i ]$ 
  & $-\sqrt{4 \over 9} [ \beta^{10}_j-\beta^{10}_i]$    \\
 & (0.3) 
  &    (0.1) \\
$[D^{**}_{1/2} \rho^+]_{{3 \over 2}P}$  
 & $+ \sqrt{16 \over 9} [  \beta^{10}_j -  \beta^{10}_i ]$ 
  & $+\sqrt{40 \over 9} [ \beta^{10}_j-\beta^{10}_i]$  \\
 & (0.3) 
  &    (1.3) \\
$[D^{**}_{1/2} \rho^+]_{{3 \over 2}F}$  
 & -
  & $0$  \\
 &  
  &    (0.0) \\
$[D^{**}_{1/2} \rho^+]_{{1 \over 2}P}$  
 & $+ \sqrt{2 \over 9} [ 5\beta^{10}_j +  \beta^{10}_i +  9\beta^{10}_k ] + \sqrt{2} \gamma^{10}$ 
  & $+\sqrt{2 \over 9} [2 \beta^{10}_j-4\beta^{10}_i -  9\beta^{10}_k] + \sqrt{2} \gamma^{10}$     \\
 & (0.8) 
  &    (5.6) \\
$[(D+D^*) a_0^+]_{{1 \over 2}P}$  
 & $+ \sqrt{2 \over 3} [ 3\beta^{01}_j -  \beta^{01}_i +  3\beta^{01}_k ] + \sqrt{2 \over 3} \gamma^{01}$ 
  & $+\sqrt{2 \over 3} [2 \beta^{01}_i+  3\beta^{01}_k] +  \sqrt{2 \over 3} \gamma^{01}$  \\
 & (2.4) 
  &    (1.0) \\
$[(D+D^*) a_1^+]_{{3 \over 2}P}$  
 & $+ \sqrt{2 \over 3} [ 4\beta^{01}_j +  \beta^{01}_i] + \sqrt{2 \over 3} \gamma^{01}$ 
  & $+\sqrt{5 \over 3} [ \beta^{01}_j-  2\beta^{01}_i] - \sqrt{1 \over 15} \gamma^{01}$  \\
 & (1.1) 
  &    (0.3) \\
$[(D+D^*) a_1^+]_{{3 \over 2}F}$  
 & -
  & $+ \sqrt{3 \over 5} \gamma^{01}$  \\
 & 
  &    (0.0) \\
$[(D+D^*) a_1^+]_{{1 \over 2}P}$  
 & $+ \sqrt{4 \over 3} [ \beta^{01}_j +  \beta^{01}_i +  3\beta^{01}_k ] + \sqrt{4 \over 3} \gamma^{01}$ 
  & $+\sqrt{4 \over 9} [2 \beta^{01}_j+2 \beta^{01}_i+  3\beta^{01}_k] +  \sqrt{4 \over 9} \gamma^{01}$  \\
 & (2.1) 
  &    (2.1) \\
$[(D+D^*) b_1^+]_{{3 \over 2}P}$  
 & $+ \sqrt{4 \over 3} [ 2\beta^{01}_j +  3\beta^{01}_i] + \sqrt{4 \over 3} \gamma^{01}$ 
  & $-\sqrt{10 \over 3}  \beta^{01}_j - \sqrt{2 \over 15} \gamma^{01}$  \\
 & (0.6) 
  &    (0.7) \\
$[(D+D^*) b_1^+]_{{3 \over 2}F}$  
 & -
  &  $+ \sqrt{6 \over 5} \gamma^{01}$  \\
 & 
  &    (0.1) \\
$[(D+D^*) b_1^+]_{{1 \over 2}P}$  
 & $+ \sqrt{2 \over 3} [ \beta^{01}_j -  3\beta^{01}_i-  3\beta^{01}_k] - \sqrt{2 \over 3} \gamma^{01}$ 
  & $-\sqrt{2 \over 3} [ 2\beta^{01}_j+ 3\beta^{01}_k] - \sqrt{2 \over 3} \gamma^{01}$  \\
 & (0.3) 
  &    (3.3) \\
$[(D+D^*) a_2^+]_{{5 \over 2}P}$  
 & -
  & $+\sqrt{12} [ \beta^{01}_j+ \beta^{01}_i] + \sqrt{48 \over 25} \gamma^{01}$  \\
 &  
  &    (2.5) \\
$[(D+D^*) a_2^+]_{{5 \over 2}F}$  
 & $+  \sqrt{16 \over 5} \gamma^{01}$ 
  & $ - \sqrt{32 \over 25} \gamma^{01}$  \\
 & (0.1) 
  &    (0.1) \\
$[(D+D^*) a_2^+]_{{3 \over 2}P}$  
 & $- \sqrt{10 \over 3} \beta^{01}_i - \sqrt{2 \over 15} \gamma^{01}$ 
  & $+\sqrt{1 \over 3} [ 3\beta^{01}_j-2\beta^{01}_i] + \sqrt{1 \over 75} \gamma^{01}$  \\
 & (0.0) 
  &    (0.6) \\
$[(D+D^*) a_2^+]_{{3 \over 2}F}$  
 & -
  &  - $\sqrt{3 \over 25} \gamma^{01}$  \\
 &  
  &    (0.0) \\
 \end{tabular}
  \label{tab:piamps}
\end{table}

\vfill\eject

\vspace{0.5cm}
\begin{table}[th]
  \caption[x]{Assessing the Uncertainties in $c_{q \bar q}$ and $\Delta \rho^2_{sea}$}
 \begin{tabular}{lccccc}
   & $\beta_{ft}$ (GeV) & $\beta_{pc}$ (GeV)$^\dagger$ & $\eta_{q \bar q}$ (GeV) & $c_{u \bar u}$ & $\Delta \rho^2_{sea}$  \\
  \hline
  canonical                  &0.60&0.58&$\sim 1$  &$\sim 1/2$&$\sim 1/4$ \\
  harder pair creation$^{(1)}$       &0.60&0.87&$\sim 3$  &$\sim 1$  &$\sim 1/2$ \\
  softer pair creation$^{(2)}$       &0.60&0.43&$\sim 1/2$&$\sim 1/3$&$\sim 1/6$ \\
  larger flux tube diameter$^{(3)}$  &0.42&0.58&$\sim 1/2$&$\sim 1/2$&$\sim 1/2$ \\
 \end{tabular}
  \label{tab:sensitivity}
\end{table}
\noindent $^\dagger$ recall $\tilde \beta_{pc} \simeq 0.7 \beta_{pc}$ so that the canonical value
of $\tilde \beta_{pc}$ is $\sim 0.4$ GeV.

\noindent $^{(1)}$    $r_q=0.3$ fm $\rightarrow$ $r_q=0.2$ fm

\noindent $^{(2)}$    $r_q=0.3$ fm $\rightarrow$ $r_q=0.4$ fm

\noindent $^{(3)}$    $f=2$ fm $\rightarrow$ $f=1$ fm

\vfill\eject

\vspace{0.5cm}
\begin{table}[th]
  \caption[x]{The fractions of the total nonresonant semileptonic decay rate predicted
in low-lying exclusive channels. These fractions are obtained from Table I by taking
$(\vert \tau_{1/2} \vert^2 + 2 \vert \tau_{3/2} \vert^2) / \Delta \rho^2_{sea}$ (which may be
obtained by summing the appropriate channels from Table I and
dividing by 3) times
flavor factors of ${3 \over 2}$ for $I=1$, ${1 \over 2}$ for $I=0$ $^{\dagger}$, and unity for $\bar K$
emission. Dominant subchannels can be read off from the single flavor rates quoted in Table I.}
  \begin{tabular}{lccc}
                       &    percent of inclusive    &  approximate phase space  &  estimated net                   \\
   channel             &     nonresonant rate in    & suppression   factor in   &  suppression after               \\
                       &    the heavy quark limit   &  $b \rightarrow c$ decays &  $\Lambda_{QCD}/m_Q$ corrections \\
  \hline
  $(D+D^*) \pi$                          &            0.9         &       1/3         &   0.5          \\
  $(D+D^*) \eta + (D+D^*) \eta'$         &            0.3         &       1/4         &   0.4          \\
  $(D+D^*) \bar K$                       &            0.6         &       1/3         &   0.5          \\
  $(D+D^*) \rho$                         &            2.8         &       1/4         &   0.4          \\
  $(D+D^*) \omega$                       &            0.9         &       1/4         &   0.4          \\
  $(D+D^*) \bar K^*$                     &            1.8         &       1/4         &   0.4          \\
  $D^{**}_{3/2} \pi$                     &            0.7         &       1/4         &   0.4          \\
  $D^{**}_{3/2} \eta+D^{**}_{3/2} \eta'$ &            0.2         &       1/5         &   0.3          \\
  $D^{**}_{s~3/2} \bar K$                &            0.5         &       1/4         &   0.4          \\
  $D^{**}_{1/2} \pi$                     &            0.7         &       1/4         &   0.4          \\
  $D^{**}_{1/2} \eta+D^{**}_{1/2} \eta'$ &            0.2         &       1/5         &   0.3          \\
  $D^{**}_{s~1/2} \bar K$                &            0.5         &       1/4         &   0.4          \\
  $D^{**}_{3/2} \rho$                    &            2.2         &       1/10        &   0.15         \\
  $D^{**}_{3/2} \omega$                  &            0.7         &       1/10        &   0.15         \\
  $D^{**}_{s~3/2} \bar K^*$              &            1.5         &       1/10        &   0.15         \\
  $D^{**}_{1/2} \rho$                    &            4.0         &       1/5         &   0.3          \\
  $D^{**}_{1/2} \omega$                  &            1.3         &       1/5         &   0.3          \\
  $D^{**}_{s~1/2} \bar K^*$              &            2.7         &       1/5         &   0.3          \\
  $(D+D^*) a_0$                          &            1.7         &       1/5         &   0.3          \\
  $(D+D^*) f_0$                          &            0.6         &       1/5         &   0.3          \\
  $(D+D^*) \bar K^*_0$                   &            1.1         &       1/5         &   0.3          \\
  $(D+D^*) a_1$                          &            2.8         &       1/6         &   0.25          \\
  $(D+D^*) f_1$                          &            0.9         &       1/6         &   0.25          \\
  $(D+D^*) \bar K_{1a}$                  &            1.9         &       1/6         &   0.25          \\
  $(D+D^*) b_1$                          &            2.4         &       1/5         &   0.3          \\
  $(D+D^*) h_1$                          &            0.8         &       1/5         &   0.3          \\
  $(D+D^*) \bar K_{1b}$                  &            1.6         &       1/5         &   0.3          \\
  $(D+D^*) a_2$                          &            1.6         &       1/6         &   0.25          \\
  $(D+D^*) f_2$                          &            0.5         &       1/6         &   0.25          \\
  $(D+D^*) \bar K^*_2$                   &            1.1         &       1/6         &   0.25          \\
  \end{tabular}
  \label{tab:rates}
\end{table}
\noindent $^\dagger$ For the separate emission of $\eta$ and $\eta'$, or consideration of OZI-violating
mixing in other nonets, mixing angles between the $\omega$- and $\phi$-like components may easily be
introduced. Note that with the neglect of OZI mixing, emission of the $\phi$-like meson is forbidden.

\end{document}